\documentclass[12pt]{article}
\pdfoutput=1
\usepackage{epsf,amsfonts,amssymb,epsfig,amsmath,graphics,cite}
\usepackage[dvipsnames]{xcolor}
\usepackage{hyperref,graphicx,subfig}
\usepackage{booktabs}
\usepackage{float}
\usepackage{verbatim}
\usepackage{mathrsfs}
\usepackage{tensor}
\usepackage{empheq}
\usepackage{enumerate}
\usepackage{bbm}
\usepackage{soul}
\usepackage{cleveref}
 
\makeatletter

\newcommand{\Rmnum}[1]{\expandafter\@slowromancap\romannumeral #1@}
\makeatother

\definecolor{revisionteal}{RGB}{0,128,128}

\newcommand{\nn}{\notag }
\def\be{\begin{equation}}
\def\ee{\end{equation}}

\newcommand{\ii}{\mathrm{i}}
\newcommand{\ex}{\mathrm{e}}
\newcommand{\diff}{\mathrm{d}}
\newcommand{\dd}{\mathrm{d}}
\newcommand{\R}{\mathbb{R}}
\newcommand{\Z}{\mathbb{Z}}
\newcommand{\vol}{\mathrm{vol}}

\newcommand{\fluxfib}{\mathfrak n}
\newcommand{\fluxpole}[1]{\mathfrak m_{#1}}

\newcommand{\hook}{\mathbin{\rule[.2ex]{.4em}{.03em}\rule[.2ex]{.03em}{.9ex}}}

\newcommand{\me}{\mathrm{e}}
\newcommand{\Fixed}{\mathscr{F}}

\numberwithin{equation}{section}       

\begin{document}

\pagenumbering{Alph}
\begin{titlepage}


\vskip 1cm

\begin{center}


{\Large \bf Localizing $\mathcal{N}=(2,0)$ $AdS_3$ solutions of M-theory}

\vskip 1cm

{Carolina Matt\'e Gregory$^{\mathrm{a,b}}$, Catherine Notman$^{\mathrm{c}}$, and James Sparks$^{\mathrm{c}}$}

\vskip 1cm

${}^{\mathrm{a}}$\textit{Universidade de Bras\'ilia, Instituto de F\'isica, \\
70910-900, Bras\'ilia, DF, Brasil }\\

\vskip 0.2cm

${}^{\mathrm{b}}$\textit{International Center of Physics, 
C.P. 04667, Bras\'ilia, DF, Brasil }\\

\vskip 0.2cm

 ${}^{\mathrm{c}}$\textit{Mathematical Institute, University of Oxford,\\
Andrew Wiles Building, Radcliffe Observatory Quarter,\\
Woodstock Road, Oxford, OX2 6GG, U.K.\\}

\vskip 0.2 cm

\end{center}

\vskip 0.5 cm

\begin{abstract}
\noindent
We apply equivariant localization to supersymmetric $AdS_3\times M_8$ solutions of eleven-dimensional supergravity, dual to two-dimensional $\mathcal N=(2,0)$ SCFTs.  We construct equivariantly closed polyforms encoding the magnetic flux and its dual, calibrated M2-branes and the warped volume, reducing protected observables to fixed point data of the $U(1)$ R-symmetry action. We apply the framework to $S^4$ bundles over coassociative four-cycles and $S^2$ bundles over compact six-manifolds, obtaining topological formulae for leading central charges and gravitational anomalies in terms of quantized fluxes and topological data. 
We recover the known results for $T^2\times KE_4$ solutions and construct a
compact numerical solution with quantized flux that lies outside the known
polynomial branch.

\end{abstract}

\end{titlepage}

\pagenumbering{arabic}

\pagestyle{plain}
\setcounter{page}{1}
\newcounter{bean}
\baselineskip18pt

\tableofcontents


\newcommand{\DA}{\lambda}
\newcommand{\chiA}{\epsilon}
\newcommand{\LA}{\xi}
\newcommand{\PhiA}{\Psi}

\section{Introduction}\label{sec:intro}

Supersymmetric $AdS_3$ solutions of string and M-theory provide a rich setting in which to study holography for two-dimensional superconformal field theories (SCFTs). A particularly important class arises from M5-branes wrapped on supersymmetric cycles, where the geometry of the wrapped cycle, flux quantization and the topology of the internal manifold encode protected data of the infra-red SCFT. Even when the complete supergravity solution is difficult to construct explicitly, quantities such as central charges, anomalies and dimensions of BPS operators are often much more rigid. It is therefore natural to ask to what extent these observables can be determined directly from supersymmetry and topology, without solving for the full supergravity solution.

Equivariant localization \cite{BenettiGenolini:2023kxp} provides a powerful framework for doing precisely this. In recent work \cite{BenettiGenolini:2023yfe,BenettiGenolini:2023ndb,Martelli:2023eq,Colombo:2023eqvol,BenettiGenolini:2024locattr,Couzens:2024fmax,BenettiGenolini:2024free,Cassani:2024loc,BenettiGenolini:2024toric,BenettiGenolini:2024d4,Couzens:2025romans,Colombo:2025odd,Suh:2024eq,BenettiGenolini:2025d5,Couzens:2025punctures,Couzens:2026ten}, supersymmetry conditions for supergravity solutions have been reorganized into equivariantly closed forms associated with the R-symmetry Killing vector, so that integrals defining holographic observables may be reduced to contributions from the fixed point set by the Berline--Vergne--Atiyah--Bott formula \cite{BerlineVergne:1982,AtiyahBott:1984}. In the $AdS_3$ context this method has been developed for particular classes of wrapped M5-brane solutions, where it reproduces flux quantization and central charges from fixed point and cohomological data alone \cite{BenettiGenolini:2023yfe,BenettiGenolini:2023ndb}. 

The purpose of this paper is to extend this framework to the general supersymmetric $AdS_3\times M_8$ solutions of eleven-dimensional supergravity studied in \cite{Ashmore:2022ydf}. These backgrounds admit a supersymmetric Killing vector $\xi$, dual to the $U(1)$ R-symmetry of the two-dimensional $\mathcal N=(2,0)$ SCFT. Starting from the spinor bilinears and differential conditions of \cite{Ashmore:2022ydf}, we construct equivariantly closed completions of the magnetic four-form, its Hodge dual and the differential forms entering calibrated M2-branes and the holographic central charge.\footnote{Whenever the scalar equivariant potentials are used globally, we assume the exactness condition \eqref{eq:global-potential-condition}.} The resulting formalism extends the constructions of \cite{BenettiGenolini:2023yfe,BenettiGenolini:2023ndb} beyond the subclass with $S=P=0$, and provides a common localization framework for a substantially broader family of $AdS_3$ solutions. 

A useful feature of the construction is that several observables are controlled by the same fixed point data. Magnetic fluxes follow from the equivariant completion of $\me^{3\lambda}F$, while the leading Brown--Henneaux central charge \cite{Brown:1986nw} is obtained from an equivariantly closed completion of the warped volume. The action of supersymmetric M2-branes is similarly governed by the calibrating two-form $\me^{3\lambda}\omega$.
The subleading difference $c_L-c_R$ of left- and right-moving central charges is determined by the gravitational Chern--Simons term \cite{Kraus_2005,KrausLarsen:2005}. In M-theory this term descends from the higher-derivative coupling $C_3\wedge X_8$  \cite{Duff:1995wd,Witten:1996hc,Freed:1998tg}.
In this case, it is important to keep track of the first Pontryagin class of the full tangent bundle $TM_8$, including contributions from the twisting of the internal fibration.

We first apply the general formalism to $S^4$ bundles over coassociative
four-cycles $B_4$, corresponding to $N$ M5-branes wrapped on coassociative
cycles in $G_2$ holonomy manifolds. An explicit supergravity solution in
this class was constructed in \cite{Gauntlett:2000ng}, and subsequently
embedded in the more general $G$-structure description of
\cite{Gauntlett:2006ux}. Independently, the anomaly analysis and
$c$-extremization of \cite{Benini:2013cda} predict, at large $N$,
\begin{equation}
 c^{(0)}=N^3\left[P_1(B_4)+2\chi(B_4)\right]
 \,,
\end{equation}
where $P_1(B_4), \chi(B_4)\in\Z$ are the Pontryagin and Euler numbers of $B_4$, respectively. 
For the explicit solution this was shown to agree with the holographic
central charge \cite{Benini:2013cda}. Here we recover the same result directly from
localization for the general $S^4$ bundle class, without requiring an
explicit form of the supergravity solution. We also obtain the
subleading gravitational anomaly\footnote{With the orientation used here, the preserved
supercharges are left-moving, so that we use an $\mathcal N=(2,0)$
convention. Reference~\cite{Benini:2013cda} instead uses
$\mathcal N=(0,2)$ with right-moving supercharges. Thus,
$(c_L,c_R)_{\rm here}=(c_R,c_L)_{\rm there}$, and hence our
$c_L-c_R$ equals their $k=c_R-c_L$.}
\begin{equation}
 c_L-c_R=\frac{N}{2}\left[P_1(B_4)+\chi(B_4)\right],
\end{equation}
as well as the conformal dimension $\Delta=2N$ of a calibrated
M2-brane wrapping the fixed two-sphere $S^2\subset S^4$ inside the four-sphere fibres.

We then turn to a much broader class of solutions in which $M_8$ is an $S^2$ bundle over a compact six-manifold $B_6$. 
Such geometries arise in the general wrapped M5-brane construction of \cite{Gauntlett:2006ux}, with infinite classes of explicit compact $S^2$ bundle solutions constructed in \cite{Gauntlett:2006qw}.
The $U(1)_R$ symmetry simply rotates the two-sphere fibres, fixing the north and south pole sections which give two copies of the base $B_6$.  The localization equations may then be formulated entirely in terms of cohomology classes, quantized magnetic fluxes and the intersection form of $B_6$. 

For the $S=P=0$ subclass, it is useful to package the magnetic fluxes into cohomology classes on the base. We denote by $\fluxfib\in H^2(B_6,\mathbb R)$ the class obtained by integrating the magnetic four-form over the $S^2$ fibre, and by $\fluxpole{\pm}\in H^4(B_6,\mathbb R)$ the flux classes obtained by restricting the magnetic four-form to the north ($+$) and south ($-$) pole sections, respectively.\footnote{The fibre periods are integral because the Pontryagin correction \cite{Witten:1996md} has vanishing fibre integral. For the projective bundle geometries used below, with $B_6$ a complex three-fold and $\mathcal L=K_{B_6}^{\pm1}$, one moreover has $p_1(TM_8)\in4H^4(M_8,\mathbb Z)$, so the Witten shift is integral and the unshifted classes $\fluxpole{\pm}$ also have integral periods, as shown in section~\ref{sec:S2B6}. Outside these hypotheses, the shifted flux class is the quantized object.} 
These classes are related by the topology of the $S^2$ bundle,
\begin{equation}
\fluxpole{+}-\fluxpole{-}=
c_1(\mathcal{L})\wedge\fluxfib\, ,
\end{equation}
so only two of the three flux classes are independent. Here 
the $S^2$ bundle is taken to be the projectivized bundle $M_8=\mathbb{P}(\mathcal O\oplus \mathcal L)$ with $\mathcal{L}$ a complex line bundle over $B_6$ that specifies the twisting of the $S^2$ fibre. The fluxes must in addition obey the compatibility condition
\begin{equation}
\int_{B_6}\fluxfib\wedge
\left(\fluxpole{+}+\fluxpole{-}\right)=0\, .
\end{equation}
We also define the map
\begin{align}
L_{\fluxfib}:H^2(B_6,\mathbb R)\longrightarrow H^4(B_6,\mathbb R)\, ,
\qquad
L_{\fluxfib}(\alpha)=\alpha\wedge\fluxfib\, .
\end{align}
When $L_{\fluxfib}$ is invertible and the denominator displayed below is
non-zero, localization then gives the leading central charge in the closed
form
\begin{align}
c^{(0)}
=
-\int_{B_6}\fluxfib^3
+
3\,
\frac{
\left(\int_{B_6}\fluxfib\wedge\fluxpole{+}\right)
\left(\int_{B_6}\fluxfib\wedge\fluxpole{-}\right)}
{
\int_{B_6}L_{\fluxfib}^{-1}(\fluxpole{+})\wedge\fluxpole{-}}\, .
\end{align}
Thus, the leading central charge is determined entirely by quantized magnetic fluxes and the intersection structure of $B_6$, without requiring the explicit supergravity solution.

The same topology gives a particularly simple expression for the gravitational anomaly. 
 We find
\begin{align}
c_L-c_R
=
-\frac14\int_{B_6}\fluxfib\wedge
\left[p_1(B_6)+c_1(\mathcal{L})^2\right]\, ,
\end{align}
where $p_1(B_6)\in H^4(B_6,\Z)$ denotes the first Pontryagin class $p_1(B_6)=p_1(TB_6)$. 
Thus, both the leading central charge and the subleading gravitational anomaly are determined by flux and topological data.

As an analytic check, we specialize these expressions to the $B_6=T^2\times KE_4$ solutions of \cite{Gauntlett:2006qw}, whose type IIB duals were studied globally in \cite{Gauntlett:2006af} and where flux quantization and the holographic central charge were first computed. The localization formula reproduces the known leading central charge, while the gravitational anomaly reduces to a purely topological expression involving Pontryagin and Euler numbers of $KE_4$.

Finally, we consider a setting where the local geometry is not known in
closed analytic form. Using the cohomogeneity one system of
\cite{Gauntlett:2006qw}, but without imposing the polynomial ansatz used to
construct the known compact families, we construct a smooth numerical
solution with $B_6=S^2\times S^2\times T^2$. The 
eleven-dimensional fields satisfy the supergravity equations numerically.
After fixing the overall scale and torus area using two magnetic flux
periods, the remaining four periods agree with integers to the numerical
precision. The resulting quantized fluxes lie outside the
canonical class ansatz of the known analytic family and violate a
necessary condition satisfied by every regular compact polynomial solution
in the ansatz. Thus the solution provides a concrete realization of the
more general flux sector described by the localization analysis.

 The remainder of the paper is organized as follows. In section~\ref{sec:general}, we review the general $AdS_3\times M_8$ supersymmetry equations and construct the relevant equivariantly closed forms, before summarizing the localization formulae for fluxes, central charges, anomalies and calibrated M2-branes. In section~\ref{sec:S4coassoc}, we study $S^4$ bundles over coassociative four-cycles. Section~\ref{sec:S2B6} develops the general $S^2$ bundle analysis and expresses the observables in terms of cohomological and intersection data. Section~\ref{sec:S2specializations} specializes these results to explicit families, including $T^2\times KE_4$ and the non-polynomial solution, before we conclude in section \ref{sec:discussion} with a discussion of the implications and open directions. We have also included several appendices: Appendix~A
derives the constraint on the magnetic flux used in the $S^2$ bundle
analysis, Appendix~B gives further details of the numerical solution, while 
Appendix~C discusses the additive freedom in the electric potential $a_0$
and its relation to the dual flux condition used in
\cite{BenettiGenolini:2023yfe}.

\section{General analysis}\label{sec:general}

Supersymmetric $AdS_3\times M_8$ solutions of eleven-dimensional supergravity dual to $d=2$, $\mathcal N=(2,0)$ SCFTs were studied in \cite{Ashmore:2022ydf}. They admit a supersymmetric Killing vector $\LA$ on $M_8$, dual to the SCFT R-symmetry. In this section, we construct polyforms on $M_8$ that are equivariantly closed under $\diff_\LA\equiv\diff-\LA\hook\mskip2mu$. This extends \cite{BenettiGenolini:2023yfe,BenettiGenolini:2023ndb}, where analogous polyforms were obtained for the subclass \cite{Gauntlett:2006qw} describing M5-branes wrapped on holomorphic four-cycles in a Calabi--Yau four-fold. The present construction applies to the more general $AdS_3$ class and hence to a broader set of supersymmetric wrapped brane configurations.

\subsection{\texorpdfstring{$AdS_3\times M_8$}{AdS3 x M8} solutions}\label{sec:M8}

We use the notation and conventions of \cite{BenettiGenolini:2023yfe,BenettiGenolini:2023ndb}, which closely follow \cite{Ashmore:2022ydf}.
The $D=11$ metric and four-form $G$ are given by
\begin{align}
\label{11dansatz}
\diff s^2_{11}&= \ex^{2\DA}\left(\dd s^2_{AdS_3}+ \dd s^2_{M_8}\right)\,,\nn\\
G&=\ex^{3\DA}F + \vol_{AdS_3}\wedge \ex^{3\DA}f\,.
\end{align}
Here $\DA$ is a function on $M_8$, while $F$ and $f$ are a four-form and a one-form, respectively. We take $\dd s^2_{AdS_3}$ to have unit radius and denote its volume form by $\vol_{AdS_3}$. The ansatz \eqref{11dansatz} is the most general one compatible with the $SO(2,2)$ symmetry of $AdS_3$. The Bianchi identity and equation of motion for $G$ are
\begin{align}\label{ads3beom}
\dd(\ex^{3\DA}F)&=0\, ,\quad  \ex^{-6\DA} \dd (\ex^{6\DA}* f)+\frac{1}{2}F\wedge F=0\,,\nn\\
\dd (\ex^{3\DA}f)&=0\, ,\quad \ \ 
\ex^{-6\DA} \dd ( \ex^{6\DA}* F)-f\wedge F=0\,.
\end{align}
Here and below, $*$ denotes the Hodge operator for $\diff s^2_{M_8}$, with the convention that a $p$-form $\alpha$ has dual $*\mskip2mu\alpha$ with components
$(*\mskip2mu\alpha)_{\mu_1\cdots\mu_{8-p}}\equiv\frac{1}{p!}\varepsilon_{\mu_1\cdots\mu_{8-p}}^{\ \ \ \ \ \ \ \ \ \ \nu_1\cdots\nu_p}\alpha_{\nu_1\cdots \nu_p}$. 

The flux of $G$ through four-cycles $\mathcal{C}_4$ satisfies the quantization condition  \cite{Witten:1996md} (see also \cite{Gauntlett:2006qw} in the present context)
\begin{align}\label{fluxquantization}
\int_{\mathcal{C}_4} \frac{1}{(2\pi\ell_p)^3}\ex^{3\DA}F -\frac{1}{4}p_1(M_8)\in \mathbb{Z}\, .
\end{align}
Here $\ell_p$ denotes the eleven-dimensional Planck length, 
and $p_1(M_8)=-\frac{1}{8\pi^2}\mathrm{Tr}\mskip2mu R\wedge R$ is the first Pontryagin class of $M_8$. For a spin manifold $M_8$, the integer class $p_1(M_8)$ is always 
divisible by two, but is not necessarily divisible by four, so that in general \eqref{fluxquantization} implies that the flux integral is half-integer. 

\subsection{Supersymmetry}\label{sec:SUSY}

Supersymmetry is encoded by a pair of Majorana spinors $\chiA_1,\chiA_2$ on $M_8$ satisfying the Killing spinor equations of \cite{Ashmore:2022ydf}. Introducing the complex spinor $\chiA\equiv\frac{1}{\sqrt{2}}(\chiA_1+\ii\chiA_2)$, define the scalar bilinears
\begin{align}
\bar\chiA\chiA=1\, ,\quad
\bar\chiA^c\chiA=0\, ,\quad
\zeta\equiv\bar\chiA\gamma_9\chiA\, ,\quad
S\equiv\bar\chiA^c\gamma_9\chiA\, .\quad
\end{align}
Here the normalization in the first equation is without loss of generality, 
while in general $\zeta$ and $S$ are real and complex functions on $M_8$, respectively. 
One can choose conventions so that the eight-dimensional gamma matrices $\gamma_\mu$ 
are real and symmetric, with $\chiA^c=\chiA^*$, $\bar\chiA=\chiA^\dagger$. 
The chirality operator  $\gamma_9 \equiv \gamma_{1} \cdots \gamma_{8}$ satisfies 
$\gamma_9^2=1$. 

Similarly, introduce the one-form bilinears\footnote{We use the same notation as 
\cite{BenettiGenolini:2023yfe, BenettiGenolini:2023ndb}, but comparing 
to  \cite{Ashmore:2022ydf} we note
$\DA=\Delta^\text{there}$, $m^\text{there}=-1/2$ (our $AdS_3$ metric has unit radius), $\chiA_i=\chi^\text{there}_i$ 
and  $\LA^\flat=\frac{1}{2}L^\text{there}$.}
\begin{align}
K\equiv\bar\chiA\gamma_{(1)}\chiA\, ,\quad
P\equiv\bar\chiA^c\gamma_{(1)}\chiA\, ,\quad
\LA^\flat\equiv-\frac{\ii}{2}\bar\chiA\gamma_9\gamma_{(1)}\chiA\, ,\quad
\bar\chiA^c\gamma_9\gamma_{(1)}\chiA=0\, ,\quad
\end{align}
where $K, \LA^\flat$ are real and $P$ is complex. We refer to the vector 
 $\LA$, dual to the one-form $\xi^\flat$, as the \emph{supersymmetric  Killing vector}. 
It was shown in \cite{Ashmore:2022ydf} that this generates a symmetry of the full solution. In particular  
$\mathcal{L}_\LA \DA=
\mathcal{L}_\LA F=\mathcal{L}_\LA f=0$, while the complex spinor $\chiA$ 
has charge $\tfrac{1}{2}$ under $\LA$:
\begin{align}\label{spinorcharged}
\mathcal{L}_\LA\chiA=\frac{\ii}{2}\chiA\, .
\end{align}

The norm of the supersymmetric Killing vector is \cite{Ashmore:2022ydf}
\begin{equation}\label{eq:xi-norm}
    \|\xi\|^2=\frac14\|K\|^2+\frac14|S|^2\,.
\end{equation}
It follows that the fixed point set of the R-symmetry action is
\begin{equation}\label{eq:fixed-locus-condition}
    \Fixed\equiv\{\xi=0\}=\{K=0,\ S=0\}\,.
\end{equation}
This criterion will be used repeatedly in the examples below.

The following real bilinear two-, three- and four-forms will also be useful
\begin{align}
J&\equiv-\ii\bar\chiA\gamma_{(2)}\chiA\,,\quad \omega\equiv-\ii\bar\chiA{\gamma_9}\gamma_{(2)}\chiA\,,\nn\\
\varphi&\equiv-\ii\bar\chiA\gamma_{(3)}\chiA\,,\quad \phi\equiv\bar\chiA{\gamma_9}\gamma_{(3)}\chiA\,,\nn\\
\PhiA&\equiv\bar\chiA\gamma_{(4)}\chiA\,,\quad\,\, * \PhiA\equiv\bar\chiA {\gamma_9}\gamma_{(4)}\chiA\,,
\end{align}
where $\gamma_{(p)}\equiv\frac{1}{p!}\gamma_{\mu_1}\cdots\gamma_{\mu_p}\diff
x^{\mu_1}\wedge\cdots\diff x^{\mu_p}$. 
Note from \eqref{spinorcharged} that the bilinears constructed from $\bar\chiA$ are all invariant under $\mathcal{L}_\LA$, while $S$ and $P$ constructed from $\bar\chiA^c$ have charge 1:  $\mathcal{L}_\LA S = 
\ii S$, $\mathcal{L}_\LA P = \ii P$. We will  make use of the following 
algebraic (Fierz) identities, relating contractions of $\LA$ into 
bilinears to other bilinears:
\begin{align}\label{Fierz}
\quad \LA\hook P & = \frac{\ii}{2}S\, , \qquad \qquad \quad \mskip3mu
\LA\hook J   = \frac{1}{2}\zeta K - \frac{1}{2}\mskip2mu \mathrm{Re}[S^*P]\, ,  \nonumber\\
 \LA \hook \omega &  = \frac{1}{2}K\, , \qquad \qquad \quad \mskip1mu
 \LA \hook \phi    = \frac{1}{2}\zeta \omega - \frac{1}{2}J + \frac{\ii}{4}P\wedge P^*\, 
, \ \nonumber\\ 
 \LA \hook \Psi  & = -\frac{1}{2}\omega\wedge K\, , \qquad \LA\hook *\mskip1mu J  = -\frac{1}{2}\Psi\wedge K
\, .
\end{align}
These may be derived straightforwardly using a computer algebra package.
Finally, we will need the following 
differential conditions on bilinears, derived in \cite{Ashmore:2022ydf}
\begin{align}\label{dbilinears}
\ex^{-3\DA} \dd (\ex^{3\DA} \zeta)&=f+2 K\,,\nn\\
\ex^{-3\DA} \dd(\ex^{3\DA} S)&= 2P\, ,\nn\\
\ex^{-3\DA} \dd (\ex^{3\DA} \LA)&= \frac{1}{2}J-\frac{1}{4}\omega\hook F+\frac{1}{4}J\hook *\mskip1mu  F\,,\nn\\
\ex^{-6\DA} \dd (\ex^{6\DA} J)&=f\wedge \omega-2\LA\hook *\mskip1mu F\,,\nn\\
\ex^{-3\DA} \dd (\ex^{3\DA} \omega)&=-2\mskip1mu \LA\hook F\,,\nn\\
\ex^{-6\DA} \dd( \ex^{6\DA} \phi) &=2\PhiA+\zeta  F-*\mskip1mu  F\,,\nn\\
\ex^{-6\DA} \dd( \ex^{6\DA }\PhiA)&=-K\wedge F\, ,\nn\\
\ex^{-9\DA} \dd(\ex^{9\DA}*J ) & = 4\mskip1mu \xi\hook \vol_8\, ,
\end{align}
where $\vol_8$ denotes the Riemannian volume form on $M_8$.

\subsection{Equivariant localization}\label{sec:equivariant-localization}

Equivariant localization provides a particularly efficient way of extracting BPS observables from supersymmetric $AdS$ solutions \cite{BenettiGenolini:2023yfe,BenettiGenolini:2023ndb}. If the R-symmetry is generated by the Killing vector $\xi$, we work in the Cartan model with $\dd_\xi=\dd-\iota_\xi$, which is nilpotent on $\xi$-invariant polyforms. Writing the fixed point set as a disjoint union of connected components $\Fixed_\alpha$, the Berline--Vergne--Atiyah--Bott formula \cite{BerlineVergne:1982,AtiyahBott:1984} for any equivariantly closed polyform $\mathcal O$ is
\begin{equation}\label{eq:general-localization}
    \int_{M_8}\mathcal O
    =\sum_\alpha \frac{1}{d_\alpha}
    \int_{\Fixed_\alpha}
    \frac{\iota_\alpha^*\mathcal O}
         {e_\xi(\mathcal N_\alpha)}\,.
\end{equation}
Here $\iota_\alpha:\Fixed_\alpha\hookrightarrow M_8$, $\mathcal N_\alpha$ is the corresponding normal bundle, $e_\xi(\mathcal N_\alpha)$ its equivariant Euler class, and $d_\alpha$ is the order of a possible orbifold structure group, and in the smooth examples below $d_\alpha=1$. For a normal complex line bundle $\mathcal{N}_\alpha$ the equivariant Euler class is normalized so that $e_\xi(\mathcal{N}_\alpha)=\frac{b}{2\pi}+c_1(\mathcal{N}_\alpha)$, where a normal two-plane is parametrized by $\varphi\sim\varphi+2\pi$ with $\xi=b\mskip2mu\partial_\varphi$, the corresponding equivariant weight is $b$, and the R-symmetry normalization fixed by \eqref{spinorcharged} allows us to set $b=1$ when appropriate. The same formula \eqref{eq:general-localization} applies after restricting to an invariant submanifold, as we will do for flux integrals. Thus, once the relevant equivariantly closed forms have been identified, the observables of interest reduce to fixed locus and topological data.

\subsection{Equivariant forms}\label{sec:equivariant}

The required equivariantly closed polyforms are constructed from the bilinears \eqref{dbilinears}, following \cite{BenettiGenolini:2023yfe,BenettiGenolini:2023ndb}.

The Bianchi identity $\dd(\ex^{3\DA}f)=0$ in \eqref{ads3beom} allows us to introduce locally a function $a_0$
via
\begin{align}\label{dda}
\ex^{3\DA} f= \dd a_0 \,. 
\end{align}
We note 
that equation~(2.49) of  \cite{Ashmore:2022ydf} gives $\LA\hook  f=0$, 
implying that $\mathcal{L}_\xi a_0=0$. 
The first equation in \eqref{dbilinears} reads
\begin{align}
\dd( \ex^{3\DA}\zeta-a_0)=2 \ex^{3\DA}K\,.
\end{align}
It follows that
$\diff (\ex^{3\DA} K)=0$, 
and we can introduce a local function $y$ via
 \begin{align}\label{ydefinition}
y\equiv\frac{1}{2}(\ex^{3\DA}\zeta-a_0)\, ,\quad K= \ex^{-3\DA} \dd y\,.
 \end{align}

Taking de Rham classes in the same scalar bilinear equation gives
\begin{equation}\label{eq:electric-global-obstruction}
 [\ex^{3\lambda}f]=-2[\ex^{3\lambda}K]
 \qquad\text{in }H^1\,.
\end{equation}
Hence smooth global choices of $a_0$ and $y$ exist precisely when
\begin{equation}\label{eq:global-potential-condition}
 [\ex^{3\lambda}K]=0
 \quad\Longleftrightarrow\quad
 [\ex^{3\lambda}f]=0\,.
\end{equation}
Thus, $H^1(M_8,\mathbb R)=0$ is sufficient, but stronger than necessary. Below we assume \eqref{eq:global-potential-condition} whenever applying localization globally.

We may construct an equivariant completion $\Phi^F$  of the four-form $F$ as
\begin{align}\label{PhiF}
\Phi^F = \Phi^F_4+\Phi^F_2+\Phi^F_0
\equiv \ex^{3\DA} F-\frac{1}{2}\ex^{3\DA}\omega-\frac{y}{4}\,.
\end{align}
To see that $\diff_\LA \Phi^F=0$, we note that $\Phi^F_4=\ex^{3\DA}F$ is closed from \eqref{ads3beom}, 
while $\LA\hook\mskip2mu \Phi^F_4 = \diff \Phi^F_2$ is the fifth equation in 
\eqref{dbilinears}. From the algebraic identity $\LA\hook \omega= \frac{1}{2}K$ in \eqref{Fierz}, 
one then checks that $\LA\hook \Phi^F_2 = \diff\Phi^F_0$ is simply \eqref{ydefinition}. 

Next, we note that defining 
 \begin{align}\label{starF}
\Phi^{*F}_4\equiv \ex^{6\DA}* F - a_0 \ex^{3\DA} F\,, 
\end{align} 
the equation of motion in \eqref{ads3beom} gives $\diff \Phi^{*F}_4=0$. This four-form equivariantly completes 
to 
\begin{align}\label{PhistarF}
\Phi^{*F}
 \equiv (\ex^{6\DA}* F - a_0 \ex^{3\DA} F) -\frac{1}{2} \left( \ex^{6\DA}J - a_0 \ex^{3\DA}\omega \right) -\frac{1}{4}y^2 + \frac{1}{16}\ex^{6\DA}|S|^2\,. 
\end{align}
The fourth and fifth equations in \eqref{dbilinears}, together with \eqref{dda}, imply 
$\LA\hook \Phi^{*F}_4 = \diff \Phi^{*F}_2$, while $\LA\hook J= \frac{1}{2}\zeta K  - \frac{1}{2}\mskip1mu \mathrm{Re}[S^*P]$ 
 from \eqref{Fierz}, together with \eqref{ydefinition} and the second equation in \eqref{dbilinears} gives 
$\LA\hook \Phi^{*F}_2 = \diff \Phi^{*F}_0$. 

There is a related equivariant form which completes the four-form $\Psi$
\begin{align}
\Phi^\Psi
&\equiv \ex^{6\lambda}\Psi + y\mskip1mu \ex^{3\lambda}F - \frac{1}{2}y\mskip1mu \ex^{3\lambda} \omega - \frac{1}{8}y^2\,.
\end{align}
Closure of $\Phi^\Psi_4$ is precisely the penultimate equation in \eqref{dbilinears} (together with \eqref{ydefinition}), 
while verifying the rest of $\diff_\LA \Phi^\Psi=0$ is similar to the computations above, where one uses the equation for $\LA\hook\Psi$ in \eqref{Fierz}. 
Using the equation for $\phi$ in \eqref{dbilinears} and 
the identity for $\LA\hook\phi$ in \eqref{Fierz}, one finds
\begin{equation}\label{eq:dual-flux-Psi-relation}
\Phi^{*F}=2\Phi^\Psi-\dd_\xi\Big(\ex^{6\lambda}\phi+\frac18\ex^{6\lambda}\mskip1mu\mathrm{Im}[S^*P]\Big)\,.
\end{equation}
The charged bilinear completion
\begin{equation}
\Phi^{\ii P\wedge P^*} \equiv \ii\mskip1mu \ex^{6\lambda} P \wedge P^* - \frac{1}{4}\ex^{6\DA}|S|^2
\end{equation}
is equivariantly exact. Consequently $[\Phi^{*F}]=2[\Phi^\Psi]$ wherever the global condition \eqref{eq:global-potential-condition} holds. The additive freedom in $a_0$ shifts the closed four-form
$\Phi^{*F}_4$ by a constant multiple of the magnetic flux. In
Appendix~\ref{app:a0-dual-flux}, we explain how this freedom is related to
the vanishing dual flux condition imposed in
\cite{BenettiGenolini:2023yfe}, namely, for a four-cycle carrying non-zero magnetic
flux, that condition fixes the additive constant in $a_0$, rather than
imposing an additional restriction on the supergravity solution.

Finally, associated with the supergravity formula for the central charge 
we have the equivariant polyform
\begin{align}\label{Phi}
\Phi 
&\equiv \ex^{9\DA}\, \vol_8 + \frac{1}{4} \ex^{9\DA} * J - \frac{1}{8}y\mskip1mu \ex^{6\DA}\PhiA - \frac{1}{16} y^2 \ex^{3\DA}F + \frac{1}{32}y^2 \ex^{3\DA}\omega + \frac{1}{192 }y^3\,. 
\end{align}
The steps for showing $\dd_\LA\Phi=0$ are identical to those above, 
where we also need the last equation in \eqref{dbilinears}. 

The equivariant forms we have constructed in this section generalize 
those of \cite{BenettiGenolini:2023yfe, BenettiGenolini:2023ndb}, 
where the latter reference assumed that the complex bilinears $S\equiv 0 \equiv P$ are identically zero, which in turn 
leads to a local ${SU}(3)$ structure. This allows us to use results from \cite{BenettiGenolini:2023yfe, BenettiGenolini:2023ndb} while extending the construction to the general class of $AdS_3$ solutions, in particular to different wrapped M5-brane configurations.

\subsection{Observables}\label{sec:observables}

We may write various physical quantities of interest in terms of integrals of the equivariantly closed polyforms we have constructed.

\subsubsection*{Fluxes}\label{sec:general-fluxes}

The magnetic flux is subject to the shifted M-theory quantization condition
\eqref{fluxquantization}. Since its equivariant completion $\Phi^F$ was
given in \eqref{PhiF}, the flux through any invariant four-cycle may be
computed by applying \eqref{eq:general-localization} to $\Phi^F$. Thus, the
later examples differ only in the fixed components, their normal weights
and Euler classes, and the topology of the relevant cycles.

\subsubsection*{Central charge}\label{sec:general-central-charge}

The integral of the top component of the equivariant form \eqref{Phi}
computes the effective Newton constant in $AdS_3$, and hence the leading
large $N$ central charge of the dual $d=2$, $\mathcal{N}=(2,0)$ SCFT:
\begin{align}\label{c}
c^{(0)} =  \frac{3}{2^5\pi^7\ell_p^9}\int_{M_8}\ex^{9\DA}\,  \vol_8
= \frac{3}{2^5\pi^7\ell_p^9}\int_{M_8}\Phi\, .
\end{align}
Because $\Phi$ is equivariantly closed, the second expression is evaluated
by \eqref{eq:general-localization}. Sections~\ref{sec:S4coassoc}--\ref{sec:S2specializations}
and the non-polynomial example then amount to evaluating the same observable on different fixed point data.

\subsubsection*{Gravitational anomaly}\label{sec:general-anomaly}

The subleading difference of left- and right-moving central charges is
determined by the gravitational Chern--Simons term \cite{KrausLarsen:2005}. Writing
\begin{equation}\label{eq:general-CS}
    S_{CS}=\beta_{CS}\int_{AdS_3}\omega_{CS}(\Gamma_{AdS_3})\, ,
    \qquad
    c_L-c_R=96\pi\beta_{CS}\, ,
\end{equation}
the relevant eleven-dimensional higher-derivative coupling \cite{Duff:1995wd, Freed:1998tg} may be written
as
\begin{align}\label{eq:general-C3X8}
    S_{CS}
    &=-\frac{(2\pi\ell_p)^6}{(2\pi)^8\ell_p^9}
      \int_{M_{11}} C_3\wedge X_8\, ,
\end{align}
where (locally) $\dd C_3=G_4$ and
\begin{equation}\label{eq:general-X8}
    X_8=
    \frac{1}{3(2\pi)^4 2^6}
    \Big[\operatorname{Tr} R^4 
    -\frac14(\operatorname{Tr}R^2)^2\Big]\, .
\end{equation}
The curvature itself admits an equivariant completion. Regarding
$\Omega^{ab}=\frac12 R^{ab}{}_{\mu\nu}\,\dd x^\mu\wedge\dd x^\nu$ as a
matrix-valued two-form, define
\begin{equation}\label{eq:PhiR}
    \Phi^R=\Omega+\mu\, ,
\end{equation}
where $\mu$ is the moment map, with components
$\mu_{ab}=-\frac12(\dd\xi)_{ab}$ \cite{BenettiGenolini:2026cdw}. For example, the
corresponding equivariant completion of the first Pontryagin form is
\begin{equation}\label{eq:Phip1}
    \Phi^{p_1}
    =-\frac{1}{8\pi^2}\operatorname{Tr}
      \bigl(\Phi^R\wedge\Phi^R\bigr)\, .
\end{equation}
Its top component is the ordinary first Pontryagin form. After reducing
\eqref{eq:general-C3X8} on the internal geometry, the anomaly is therefore
expressed in terms of the magnetic flux and Pontryagin classes; i.e., whenever
localization applies these factors are replaced by $\Phi^F$ and
$\Phi^{p_1}$ and evaluated using \eqref{eq:general-localization}. 

In the flux normalization and sign convention used throughout this paper, the reduction may be summarized as
\begin{equation}\label{eq:anomaly-cohomological}
    c_L-c_R 
    =-\frac{1}{4}\,
      \frac{1}{(2\pi\ell_p)^3}
      \int_{M_8}\me^{3\lambda}F\wedge p_1(TM_8)\,.
\end{equation}
Here the orientation on $M_8$ is determined by $\vol_8$. When evaluating this integral for the bundle constructions that appear later, where we integrate over the fibres, this orientation must be compared with the
product of the chosen base and fibre orientations.

\subsubsection*{Conformal dimensions}\label{sec:general-conformal-dimensions}

For supersymmetric M2-branes calibrated by $\ex^{3\lambda}\omega$, as in
\cite{BenettiGenolini:2023ndb}, a calibrated two-cycle
$\Sigma_2\subset M_8$ gives a chiral primary whose conformal dimension is
\begin{equation}\label{eq:DeltaM2}
    \Delta(\Sigma_2)
    =\frac{1}{(2\pi)^2\ell_p^3}
      \int_{\Sigma_2}\ex^{3\lambda}\omega\, .
\end{equation}
The remaining task in each family is therefore to identify the relevant
supersymmetric two-cycles and evaluate this universal expression.

\section{$S^4$ bundles over coassociative four-cycles}\label{sec:S4coassoc}

In this section, we consider a class of $AdS_3$ backgrounds arising from M5-branes wrapped on a coassociative four-manifold $B_4$ inside a $G_2$ holonomy manifold. 
This is the class discussed in section~4.5 of \cite{Ashmore:2022ydf}, with a particular explicit solution constructed in section~9 of \cite{Gauntlett:2006ux}. 
Our analysis is general, and in particular includes this solution.

\subsection{Geometry and fixed loci}
The coassociative class of \cite{Ashmore:2022ydf} is characterized by
\begin{equation}
\zeta=0\, ,\qquad K=0\, ,\qquad f=0\, ,
\end{equation}
while the complex scalar bilinear may be written locally as
\begin{equation}
S=\me^{\ii\sigma}\sin\alpha\, .
\end{equation}
In particular, using the definition \eqref{ydefinition},
$K=\me^{-3\lambda}\diff y$,
we see that $y$ is constant. The freedom to shift the potential $a_0$ by a constant allows us to set
$y=0$ 
without loss of generality. Equivalently, since $f=0$ and $\zeta=0$, one may choose the constant $a_0$ to vanish.

This considerably simplifies the equivariant forms of section~\ref{sec:equivariant}. In particular,
\begin{equation}
\Phi^F
=
\me^{3\lambda}F
-\frac12\me^{3\lambda}\omega\, ,
\end{equation}
while the equivariant form computing the leading central charge becomes
\begin{equation}
\Phi
=
\me^{9\lambda}\vol_8
+\frac14\me^{9\lambda}*J\, .
\end{equation}
It follows immediately from the localization formula that only six-dimensional components of the fixed point set can contribute to $c^{(0)}$. Similarly, on any invariant four-cycle on which the $U(1)_R$ action is non-trivial, the flux may receive contributions only from two-dimensional components of the fixed point set. These observations will account for the simple localization formulae below.
The fixed point set itself also follows directly from the general bilinear identities. Since $K=0$, equation~\eqref{eq:xi-norm} reduces to
$\|\xi\|^2=\frac14|S|^2$, 
and hence
\begin{equation}
\Fixed
=
\{\xi=0\}
=
\{S=0\}
=
\{\sin\alpha=0\}\, .
\end{equation}

We now describe the topology of the internal space. For M5-branes wrapping a coassociative four-manifold $B_4$ inside a $G_2$ holonomy manifold $Y_7$, the rank three normal bundle of $B_4$ in $Y_7$ is naturally identified with the bundle of self-dual two-forms \cite{McLean:1998},
\begin{equation}
N_{B_4/Y_7}=\Lambda_+^2T^*B_4\, .
\end{equation}
The five transverse directions to the wrapped M5-branes therefore form the rank five bundle
\begin{equation}\label{Esplit}
E
=
\Lambda_+^2T^*B_4\oplus\mathbb{R}_R^2\, ,
\end{equation}
where $\mathbb R_R^2$ denotes the trivial oriented two-plane on which  $U(1)_R$ acts by rotations. The internal space has the topology of the corresponding unit sphere bundle
\begin{equation}
S^4\hookrightarrow M_8=S(E)\stackrel{\varpi}{\longrightarrow} B_4\, ,
\end{equation}
as in the wrapped M5-brane geometries of \cite{Gauntlett:2006ux}, $\varpi$ denotes the projection map.

There is also an interesting way to see why the $U(1)_R$ action must have a two-dimensional fixed locus inside each $S^4$ fibre. The four-form flux through the fibre is non-zero, while for $y=0$ the equivariant completion $\Phi^F$ has only four-form and two-form components. Localization of the fibre flux therefore requires a two-dimensional fixed component. Geometrically, the $U(1)_R$ action rotates the $\mathbb R_R^2$ factor of \eqref{Esplit}
and acts trivially on the $\Lambda_+^2T^*B_4$ directions. Its fixed locus in each $S^4$ fibre is thus the unit sphere in the rank three bundle $\Lambda_+^2T^*B_4$, namely an $S^2$. Globally,
\begin{equation}\label{eq:S4-fixed-bundle}
S^2\hookrightarrow \Fixed=\Fixed_6
=S\left(\Lambda_+^2T^*B_4\right)
\stackrel{\pi}{\longrightarrow} B_4\, .
\end{equation}
Thus, the fixed point set relevant for the central charge is a six-dimensional $S^2$ bundle over the wrapped coassociative four-manifold.

We conclude this subsection with two technical comments on orientation conventions and flux quantization. First, we fix the orientations using the local $SU(3)$ frame of
\cite{Ashmore:2022ydf}. At $S=0$, taking $\cos\alpha=1$, the 
bilinears and volume form obey
\begin{align}
 P=e^8+\ii e^7,\qquad J=-e^{78},\qquad
 \omega\big|_{\Fixed}=-j,\qquad
 \vol_8=\frac16j^3\wedge e^{78}.
\end{align}
Since $\diff S=2P$ at the fixed set and $\mathcal L_\xi S=\ii S$,
$S$ is a complex normal coordinate of weight $+1$. Its complex orientation
is therefore represented by
\begin{align}
 \frac{\ii}{2}\diff S\wedge\diff\bar S=4J=-4e^{78}.
\end{align}
We use the coassociative orientation on $B_4$ and orient the fixed $S^2$
fibres so that $j$ is positive on those fibres. The horizontal part of $j$ is
the self-dual two-form selected by the point of
$S(\Lambda_+^2T^*B_4)$, whose square is positive in the base orientation.
The resulting sphere bundle orientation of $\Fixed$ is therefore $j^3/6$.
Now orient the $S^4$ fibres compatibly with these $S^2$ fibres and the
complex normal plane with coordinate $S$. Writing ``bun'' for the resulting bundle
orientations, the induced localization orientation and the supergravity
orientation then satisfy
\begin{align}
 [\Fixed]_{\mathrm{loc}}=-[\Fixed]_{\mathrm{bun}},\qquad
 [M_8]_{\mathrm{sugra}}=-[M_8]_{\mathrm{bun}}.
\end{align}
All characteristic numbers of $B_4$ below use its coassociative
orientation, while fibre integrals and sphere bundle integrations use
the bundle orientations just specified. 

Secondly,  it is also useful to note that there is no half-integral Witten shift in the flux quantization condition \eqref{fluxquantization} for this class of manifolds. Indeed, using the same Wu class argument as in \cite{GukovSparks:2001}, one finds that $p_1(TM_8)$ is divisible by four. Thus, $\frac14p_1(TM_8)$ is an integral class, and the shift may be absorbed into the integer flux labels.

\subsection{Fluxes}
The near-horizon limit of $N$ M5-branes wrapped on the coassociative four-manifold $B_4$ gives the $S^4$ bundle geometry described above, with $N$ units of magnetic four-form flux through the $S^4$ fibre \cite{Gauntlett:2006ux}. We take $N>0$ with the bundle orientation specified above.  Thus,
\begin{equation}
N=\frac{1}{(2\pi\ell_p)^3}\int_{S^4}\me^{3\lambda}F\, .
\end{equation}
Localizing this integral using $\Phi^F$ gives
\begin{equation}
N=-\frac{1}{(2\pi\ell_p)^3}\pi\int_{S^2}\me^{3\lambda}\omega\mskip1mu\big|_{\Fixed}\, ,
\end{equation}
where we have used the R-symmetry normalization fixed in section~\ref{sec:equivariant-localization}. Since $\diff(\me^{3\lambda}\omega)=-2\me^{3\lambda}\xi\hook F$, its restriction to $\Fixed$ is closed. We therefore define
\begin{equation}
\beta=[\me^{3\lambda}\omega\mskip1mu |_{\Fixed}]\in H^2(\Fixed,\mathbb R)\, ,
\end{equation}
whose period over the $S^2$ fibre is
\begin{equation}
\int_{S^2}\beta=-8\pi^2\ell_p^3N\, .
\end{equation}

Let $e\in H^2(\Fixed,\mathbb R)$ denote the Euler class of the $S^2$ fibration, normalized so that $\int_{S^2}e=2$. We may then write
\begin{equation}\label{eq:S4-beta-decomposition}
\beta=-4\pi^2\ell_p^3N\mskip1mu e+\pi^*\beta_\perp\, ,
\qquad
\beta_\perp\in H^2(B_4,\mathbb R)\, .
\end{equation}
We now show that $\beta_\perp=0$.
In this section $f=K=0$, so the global condition \eqref{eq:global-potential-condition} is automatically satisfied. Restricting the cohomological relation \eqref{eq:dual-flux-Psi-relation} to the fixed point set, and using $a_0=y=0$, therefore gives
\begin{equation}
[\me^{6\lambda}*F]=2[\me^{6\lambda}\Psi]=-\beta^2\, ,
\end{equation}
where in the last equality we used the coassociative parameterization of \cite{Ashmore:2022ydf}. 
Let $\Sigma_2\subset B_4$ be any two-cycle, and let $\mathcal C_4$ denote the restriction of the fixed $S^2$ bundle to $\Sigma_2$. This four-cycle bounds a $B^3$ bundle over $\Sigma_2$ inside the corresponding $S^4$ bundle: at fixed azimuthal angle on $\R^2$ the $S^4\subset\mathbb R^3\oplus\mathbb R^2$ fibre reduces to a three-ball whose boundary is the fixed $S^2$.  Since $f=0$, the flux equation gives $\diff(\me^{6\lambda}*F)=0$, and hence Stokes' theorem implies
\begin{equation}
\int_{\mathcal C_4}\beta^2=0\, .
\end{equation}
Using \eqref{eq:S4-beta-decomposition}, only the cross term contributes since the 
Bott--Cattaneo formula \cite{BottCattaneo:1998} gives $\pi_*(e^2)=0$ where $\pi$ is the projection \eqref{eq:S4-fixed-bundle}, thus giving
\begin{equation}
0=-16\pi^2\ell_p^3N\int_{\Sigma_2}\beta_\perp\, .
\end{equation}
Since $\Sigma_2$ is arbitrary, all periods of $\beta_\perp$ vanish, and therefore
$\beta_\perp=0$. 
Consequently, the fixed point flux class is completely determined by the M5-brane charge,
\begin{equation}\label{eq:S4-beta-final}
\beta=-4\pi^2\ell_p^3N \mskip1mu e \, .
\end{equation}

\subsection{Central charge}
We now apply the universal central charge formula \eqref{c}. Using the parameterization of the coassociative class in section~4.5 of \cite{Ashmore:2022ydf}, on the fixed point set one has
\begin{equation}
*J\big|_{\Fixed}=\frac{1}{6}\omega^3\big|_{\Fixed}\, .
\end{equation}
The localization formula therefore gives, with the induced fixed point set
orientation displayed explicitly,
\begin{equation}\label{eq:S4-central-localized}
c^{(0)}=\frac{\pi}{(2\pi)^7\ell_p^9}
\int_{\Fixed_{\mathrm{loc}}}\beta^3\, .
\end{equation}

The remaining integral is topological. As shown in section~3.1, the fixed point set is the $S^2$ bundle $\Fixed=S(\Lambda_+^2T^*B_4)$ \eqref{eq:S4-fixed-bundle}. For the sphere bundle $S(V)$ of an oriented rank three vector bundle $V$, the Bott--Cattaneo formula \cite{BottCattaneo:1998} gives
\begin{equation}\label{eq:Bott-Cattaneo}
\pi_*(e^3)=2p_1(V)\, ,
\end{equation}
where $e$ is the Euler class of the vertical tangent bundle, normalized as in section~3.2. This push-forward/fibre integration uses the bundle-oriented $S^2$ fibres. For
$V=\Lambda_+^2T^*B_4$, substituting \eqref{eq:S4-beta-final} and using
$[\Fixed]_{\mathrm{loc}}=-[\Fixed]_{\mathrm{bun}}$ therefore gives
\begin{equation}\label{eq:S4-leading-central-charge}
\begin{aligned}
c^{(0)}
&=-\frac{N^3}{2}\int_{\Fixed_{\mathrm{loc}}}e^3
 =\frac{N^3}{2}\int_{\Fixed_{\mathrm{bun}}}e^3\\
&=N^3\int_{B_4}p_1(\Lambda_+^2T^*B_4)\, .
\end{aligned}
\end{equation}
For an oriented four-manifold,
\begin{equation}\label{eq:S4-normal-p1}
p_1(\Lambda_+^2T^*B_4)=p_1(TB_4)+2e(TB_4)\, ,
\end{equation}
and hence
\begin{equation}\label{eq:S4-leading-topological}
c^{(0)}=N^3\left[P_1(B_4)+2\chi(B_4)\right]\, ,
\end{equation}
in agreement with the large $N$ result of \cite{Benini:2013cda}. Using the Hirzebruch signature theorem, this may equivalently be written in terms of the signature $\sigma(B_4)$ as
\begin{equation}
c^{(0)}=N^3\left[3\sigma(B_4)+2\chi(B_4)\right]\, .
\end{equation}
In particular, positivity of the holographic central charge requires $3\sigma(B_4)+2\chi(B_4)>0$ for any regular solution in this class.
\subsection{Gravitational anomaly}
The gravitational anomaly follows directly from the topology of the $S^4$ bundle. From the decomposition of the rank five bundle $E$ in \eqref{Esplit}, together with $M_8=S(E)$, the tangent bundle satisfies
\begin{equation}
TM_8\oplus\mathbb R\simeq\varpi^*\left(TB_4\oplus\Lambda_+^2T^*B_4\oplus\mathbb R_R^2\right)\, ,
\end{equation}
where here $\varpi:M_8\rightarrow B_4$ denotes the $S^4$ bundle projection. It follows that
\begin{equation}
p_1(TM_8)=\varpi^*\left[p_1(TB_4)+p_1(\Lambda_+^2T^*B_4)\right]\, .
\end{equation}
Using \eqref{eq:S4-normal-p1}, we therefore find
\begin{equation}\label{eq:S4-full-p1}
p_1(TM_8)=2\varpi^*\left[p_1(TB_4)+e(TB_4)\right]\, .
\end{equation}
The same orientation change is required here. Since
$[M_8]_{\mathrm{sugra}}=-[M_8]_{\mathrm{bun}}$ and $N$ is the flux
through the bundle-oriented $S^4$ fibre, the integral is
\begin{align}
\begin{aligned}
\frac{1}{(2\pi\ell_p)^3}
\int_{(M_8)_{\mathrm{sugra}}}\me^{3\lambda}F\wedge p_1(TM_8)
 &=-N\int_{B_4}\left[p_1(TB_4)+p_1(\Lambda_+^2T^*B_4)\right]\\
 &=-2N\left[P_1(B_4)+\chi(B_4)\right].
\end{aligned}
\end{align}
Substituting into \eqref{eq:anomaly-cohomological} gives
\begin{equation}\label{eq:S4-anomaly}
c_L-c_R=\frac{N}{2}\left[P_1(B_4)+\chi(B_4)\right]\, .
\end{equation}
This agrees with the $O(N)$ anomaly in \cite{Benini:2013cda}
after matching chirality conventions. Their supersymmetric sector is
right-moving, while ours is left-moving, so that
$(c_L,c_R)_{\mathrm{here}}=(c_R,c_L)_{\mathrm{there}}$ and hence
$(c_L-c_R)_{\mathrm{here}}=k_{\mathrm{there}}$.

\subsection{Calibrated M2-branes}
Finally, the fixed two-sphere $S^2\subset\Fixed$ gives a natural calibrated cycle for a supersymmetric M2-brane.
Since $\int_{S^2}\beta<0$ in the bundle orientation, its calibrated
orientation is $\Sigma_2=-S^2$. Using \eqref{eq:DeltaM2} and the flux
period determined in section~3.2, its conformal dimension is
\begin{equation}
\Delta
=\frac{1}{(2\pi)^2\ell_p^3}\int_{\Sigma_2}\beta
=-\frac{1}{(2\pi)^2\ell_p^3}\int_{S^2}\beta
=2N\, .
\end{equation}
Thus, this class of solutions contains a universal chiral primary of dimension $2N$, independent of the topology of $B_4$ and of the explicit form of the supergravity solution.

\section{$S^2$ bundles over six-manifolds}\label{sec:S2B6}
In this section, we consider a general class of solutions in which the internal manifold $M_8$ is an $S^2$ bundle over a compact six-manifold $B_6$, with the $U(1)_R$ symmetry acting by rotations of the fibre. This topology arises naturally in the wrapped M5-brane construction of \cite{Gauntlett:2006ux}, and particular infinite families of explicit compact solutions were constructed in \cite{Gauntlett:2006qw}, including the solution of \cite{Gauntlett:2000ng}.

Our aim here is to treat this topology directly, without assuming an explicit form of the supergravity solution. Using equivariant localization, we derive a general expression for the leading central charge in terms of quantized magnetic fluxes and topological data on $B_6$. The previously known explicit solutions then arise as special cases of this general formula.

\subsection{Geometry and fluxes}
We consider the $S^2$ bundle
\begin{equation}\label{eq:S2-bundle}
S^2\hookrightarrow
M_8=\mathbb P(\mathcal O\oplus\mathcal{L})
\stackrel{\pi}{\longrightarrow} B_6\, ,
\end{equation}
where $\mathcal{L}\rightarrow B_6$ is a complex line bundle. Equivalently, $M_8$ is the unit sphere bundle of the oriented rank three real bundle
$\mathbb R\oplus\mathcal{L}_{\mathbb R}$. The $U(1)_R$ symmetry acts by rotations of $\mathcal{L}_{\mathbb R}$ and has two fixed components,
\begin{equation}\label{eq:S2-fixed-set}
\Fixed=B_6^+\sqcup B_6^-\, ,
\end{equation}
given by the north and south pole sections of the $S^2$ bundle. We denote the corresponding inclusions by
$\iota_\pm:B_6^\pm\hookrightarrow M_8$. Their normal bundles and equivariant weights are
\begin{equation}
\mathcal N_+\simeq\mathcal{L}\, ,\quad
\mathcal N_-\simeq\mathcal{L}^{-1}\, ,\quad
\epsilon_+=1\, ,\quad
\epsilon_-=-1\, ,
\end{equation}
where we have used the R-symmetry normalization fixed in section~\ref{sec:equivariant-localization}. We use the projective bundle orientation on $M_8$, together with the chosen orientation of $B_6$. With these conventions, the two fixed sections inherit the orientation of $B_6$, and the relative signs in the localization formula are encoded by the weights $\epsilon_\pm=\pm1$.

The global potentials introduced in section~\ref{sec:equivariant} exist for this class of geometries. Indeed, since the $S^2$ fibre is simply connected,
\begin{equation}
\pi^*:H^1(B_6,\mathbb R)\longrightarrow H^1(M_8,\mathbb R)
\end{equation}
is an isomorphism. Since $\iota_\pm$ are sections of $\pi$, the pull-back of $\me^{3\lambda}K$ to either fixed component is zero by \eqref{eq:fixed-locus-condition}, and hence
$[\me^{3\lambda}K]=0$. Equation~\eqref{eq:electric-global-obstruction} then also gives
\begin{equation}\label{eq:S2-global-potentials}
[\me^{3\lambda}K]=[\me^{3\lambda}f]=0\, .
\end{equation}
Thus, global functions $a_0$ and $y$ required for the equivariant construction may be chosen
from the outset.

We now introduce the magnetic flux data. The flux through the $S^2$ fibre defines the class
\begin{equation}\label{eq:S2-fibre-flux-class}
\fluxfib
\equiv
\pi_*\left[
\frac{\me^{3\lambda}F}{(2\pi\ell_p)^3}
\right]
\in H^2(B_6,\mathbb R)\, ,
\end{equation}
while restriction to the two fixed sections defines
\begin{equation}\label{eq:S2-pole-flux-classes}
\fluxpole{\pm}
\equiv
\iota_\pm^*\left[
\frac{\me^{3\lambda}F}{(2\pi\ell_p)^3}
\right]
\in H^4(B_6,\mathbb R)\, .
\end{equation}
Thus, the periods of $\fluxfib$ are the magnetic fluxes through the $S^2$ bundles over two-cycles in $B_6$, while the periods of $\fluxpole{\pm}$ are the magnetic fluxes through four-cycles in the two pole sections $B_6^\pm$.

These three classes are not independent. Let
$u\in H^2(M_8,\mathbb Z)$ denote the standard degree two class of the projective bundle (see e.g. \cite{BottTu}), normalized so that
\begin{equation}\label{bob}
\pi_*u=1\, ,\quad
\iota_-^*u=0\, ,\quad
\iota_+^*u=c_1(\mathcal{L})\, .
\end{equation}
Equivalently, the projective bundle relation is
\begin{equation}
u^2 = u\wedge \pi^*c_1(\mathcal L)\, .
\end{equation}
Here the first equation in \eqref{bob} may also be written $\int_{S^2} u = 1$, where $u$ may be represented by a closed two-form which integrates to 1 through the $S^2$ fibres. 
Since $u$ restricts to a generator of $H^2(S^2,\mathbb Z)$ on each fibre, the Leray--Hirsch theorem implies that every degree four cohomology class on $M_8$ may be written uniquely as the sum of a pull-back from $B_6$ and $u$ wedged with the pull-back of a degree two class \cite{BottTu}. Applying this to the magnetic flux class gives
\begin{equation}
\left[
\frac{\me^{3\lambda}F}{(2\pi\ell_p)^3}
\right]
=
\pi^*\fluxpole{-}
+u\wedge\pi^*\fluxfib\, .
\end{equation}
Restricting this expression to the two fixed sections immediately gives
\begin{equation}\label{eq:S2-flux-relation}
\fluxpole{+}-\fluxpole{-}
=
c_1(\mathcal{L})\wedge\fluxfib\, .
\end{equation}
Thus, once the fibre flux and the flux on one pole section are specified, the flux on the other section is fixed by the topology of the $S^2$ bundle.

Finally, we briefly discuss flux quantization. The M-theory condition \eqref{fluxquantization} involves the Pontryagin class of the full tangent bundle. The tangent bundle of \eqref{eq:S2-bundle} gives
\begin{equation}\label{eq:S2-p1-general}
p_1(TM_8)
=
\pi^*\left[p_1(B_6)+c_1(\mathcal{L})^2\right]\, .
\end{equation}
In particular, the Pontryagin correction has vanishing fibre integral, so the periods of $\fluxfib$ are integral.

For the geometries relevant to the explicit solutions below, $B_6$ is a complex three-fold and $\mathcal{L}=K_{B_6}$ with the latter the canonical complex line bundle, i.e., the same argument applies to $\mathcal{L}=K_{B_6}^{-1}$. In this case
\begin{equation}
p_1(TM_8)
=
2\pi^*\left[c_1(B_6)^2-c_2(B_6)\right]\, .
\end{equation}
The Wu relation on an oriented six-manifold gives $w_4=w_2^2$, while for a complex manifold $w_2=c_1\bmod2$ and $w_4=c_2\bmod2$. Hence $c_1(B_6)^2-c_2(B_6)$ is divisible by two, and therefore
\begin{equation}
p_1(TM_8)\in4H^4(M_8,\mathbb Z)\, .
\end{equation}
Thus, there is no half-integral Witten shift in this case, and the unshifted classes $\fluxfib$ and $\fluxpole{\pm}$ themselves have integral periods. This is the same Wu class argument used in \cite{GukovSparks:2001}.

\subsection{Central charge}\label{subsec:S2-central-charge}
We now restrict to the $S=P=0$ subclass. As reviewed in
\cite{BenettiGenolini:2023ndb,Ashmore:2022ydf}, in this case the scalar
bilinear may be written
\begin{equation}
\zeta=\sin\alpha\, ,
\end{equation}
and the bilinears introduced in section~\ref{sec:SUSY} may be parameterized
in terms of a local $SU(3)$ structure on the six-dimensional distribution
orthogonal to two orthonormal one-forms $e^7,e^8$. Denoting the corresponding real
two-form by $j$, one has
\begin{align}\label{eq:S2-SU3-parametrization}
\xi^\flat&=\frac12\cos\alpha\,e^7\, ,\quad
K=\cos\alpha\,e^8\, ,\nonumber\\
J&=j+\sin\alpha\,e^{78}\, ,\quad
\omega=\sin\alpha\,j+e^{78}\, ,\nonumber\\
\Psi&=-\frac12j^2-\sin\alpha\,j\wedge e^{78}\, ,
\end{align}
where $e^{78}\equiv e^7\wedge e^8$.

Since the fixed point set is $\Fixed=\{\xi=0\}$, equation
\eqref{eq:S2-SU3-parametrization} implies that $\cos\alpha=0$ on the two
fixed sections $B_6^\pm$. Choosing the signs to agree with the north and
south pole conventions of section~4.1, we have
\begin{equation}
\sin\alpha\big|_{B_6^\pm}=\pm1\, .
\end{equation}
Pulling back \eqref{eq:S2-SU3-parametrization} to the fixed sections then
gives
\begin{equation}\label{eq:S2-fixed-SU3-data}
\omega\big|_{B_6^\pm}=\pm j_\pm\, ,\quad
\Psi\big|_{B_6^\pm}=-\frac12j_\pm^2
=-\frac12\omega^2\big|_{B_6^\pm}\, ,\quad
(*J)\big|_{B_6^\pm}
=\pm\frac16j_\pm^3
=\frac16\omega^3\big|_{B_6^\pm}\, .
\end{equation}
Moreover, the differential condition for $\omega$ in \eqref{dbilinears} implies that $\me^{3\lambda}\omega$ is closed after restriction to either fixed section. We therefore define
\begin{equation}\label{eq:S2-beta-def}
\beta_\pm\equiv
\Big[\me^{3\lambda}\omega\big|_{B_6^\pm}\Big]
\in H^2(B_6,\mathbb R)\, ,
\end{equation}
and write $y_\pm\equiv y|_{B_6^\pm}$. Since $\diff y=\me^{3\lambda}K$ by \eqref{ydefinition}, and $K=0$ on the fixed set by \eqref{eq:fixed-locus-condition}, $y_\pm$ are constant on each connected component. The common additive freedom in $y$ allows us to choose
\begin{equation}\label{eq:S2-y-gauge}
y_-=0\, .
\end{equation}
It is then convenient to set
\begin{equation}\label{eq:S2-xs-def}
\Lambda\equiv(2\pi\ell_p)^3\, ,\quad
x\equiv-\frac{\pi}{\Lambda}\beta_-\in H^2(B_6,\mathbb R)\, ,\quad
s\equiv\frac{\pi^2}{\Lambda}y_+\in\mathbb R\, .
\end{equation}

We first relate these fixed point data to the magnetic fluxes. Let
$\Sigma_2\subset B_6$ be an arbitrary two-cycle, and denote by
$\widehat\Sigma_4=\pi^{-1}(\Sigma_2)$ the corresponding $S^2$ bundle.
Its fixed point set consists of the two copies
$\Sigma_2^\pm\subset B_6^\pm$. From \eqref{PhiF}, the components of the
equivariant flux polyform that contribute on these two-dimensional fixed
loci are
\begin{equation}
\Phi^F_2\big|_{B_6^\pm}=-\frac12\beta_\pm\, ,\quad
\Phi^F_0\big|_{B_6^\pm}=-\frac14y_\pm\, .
\end{equation}
For a normal complex line with equivariant weight
$\epsilon_\pm=\pm1$, the inverse equivariant Euler class entering
\eqref{eq:general-localization} is
\begin{equation}
\frac{1}{e_\xi(\mathcal N_\pm)}
=
\frac{2\pi}{\epsilon_\pm}
\bigg[
1-\frac{2\pi}{\epsilon_\pm}c_1(\mathcal N_\pm)
\bigg]+\text{terms of degree greater than two}\, .
\end{equation}
Using
$c_1(\mathcal N_\pm)=\pm c_1(\mathcal L)$, localization therefore gives
\begin{align}
\int_{\Sigma_2}\fluxfib
&=
\frac{1}{\Lambda}
\int_{\widehat\Sigma_4}\me^{3\lambda}F
\nn\\
&=
\int_{\Sigma_2}
\left[
-\frac{\pi}{\Lambda}(\beta_+-\beta_-)
+\frac{\pi^2}{\Lambda}(y_+-y_-)\,c_1(\mathcal L)
\right]\, .
\end{align}
Since this holds for every two-cycle $\Sigma_2$, and using the gauge choice
$y_-=0$ together with \eqref{eq:S2-xs-def}, we obtain the cohomological
relation
\begin{equation}\label{eq:S2-beta-flux-relation}
\fluxfib
=
-\frac{\pi}{\Lambda}(\beta_+-\beta_-)
+s\,c_1(\mathcal L)\, .
\end{equation}
Thus,
\begin{equation}\label{eq:S2-beta-xs}
-\frac{\pi}{\Lambda}\beta_-=x\, ,\quad
-\frac{\pi}{\Lambda}\beta_+
=x+\fluxfib-s\,c_1(\mathcal L)\, .
\end{equation}
The corresponding localization of $\Phi^F$ on $S^2$ bundles over four-cycles gives precisely the topological relation \eqref{eq:S2-flux-relation}, so introduces no further independent constraint.

We may now localize the universal central charge polyform \eqref{Phi}. Using
\eqref{eq:S2-fixed-SU3-data}, its non-zero components after restriction to
either fixed section are, in cohomology,
\begin{align}\label{eq:S2-Phi-fixed-components}
\Phi_6\big|_\pm&=\frac{1}{24}\beta_\pm^3\, ,\quad
\Phi_4\big|_\pm=\frac{1}{16}y_\pm\beta_\pm^2
-\frac{\Lambda}{16}y_\pm^2\fluxpole{\pm}\, ,\nonumber\\
\Phi_2\big|_\pm&=\frac{1}{32}y_\pm^2\beta_\pm\, ,\quad
\Phi_0\big|_\pm=\frac{1}{192}y_\pm^3\, .
\end{align}
To obtain a six-form on $B_6^\pm$ in the BVAB formula, we expand the inverse
equivariant Euler class to cubic order. Since $\epsilon_\pm=\pm1$ and
$c_1(\mathcal N_\pm)=\pm c_1(\mathcal L)$, the ratio
$c_1(\mathcal N_\pm)/\epsilon_\pm$ is the same at both poles. Thus,
\begin{equation}
\frac{1}{e_\xi(\mathcal N_\pm)}
=
\frac{2\pi}{\epsilon_\pm}
\sum_{k=0}^3
\left[-2\pi c_1(\mathcal L)\right]^k
+\text{terms of degree greater than six}\, .
\end{equation}
It follows from \eqref{c} and \eqref{eq:general-localization} that
\begin{equation}\label{eq:S2-central-localized-intermediate}
c^{(0)}
=
\frac{3}{2^5\pi^7\ell_p^9}\,2\pi
\int_{B_6}\left(\mathcal I_+-\mathcal I_-\right)\, ,
\end{equation}
where
\begin{equation}
\mathcal I_\pm
=
\frac{1}{24}
\left[\beta_\pm-\pi y_\pm c_1(\mathcal L)\right]^3
+\frac{\pi\Lambda}{8}\,
y_\pm^2c_1(\mathcal L)\wedge\fluxpole{\pm}\, .
\end{equation}
Using $y_-=0$ together with \eqref{eq:S2-beta-xs}, we have
\begin{equation}
\beta_-=-\frac{\Lambda}{\pi}x\, ,\qquad
\beta_+-\pi y_+c_1(\mathcal L)
=-\frac{\Lambda}{\pi}(x+\fluxfib)\, .
\end{equation}
Substituting these expressions into
\eqref{eq:S2-central-localized-intermediate}, and using
$\Lambda=(2\pi\ell_p)^3$, gives
\begin{equation}\label{eq:S2-central-xs}
c^{(0)}
=-4\int_{B_6}
\left[
(x+\fluxfib)^3-x^3
-3s^2c_1(\mathcal L)\wedge\fluxpole{+}
\right]\, .
\end{equation}
The remaining task is to eliminate the cohomology class $x$ and parameter $s$ in favour of the magnetic flux data.

A second cohomological relation follows by localizing the equivariantly
closed polyform $\Phi^\Psi$, defined immediately above
\eqref{eq:dual-flux-Psi-relation}, on the $S^2$ bundle over an arbitrary
four-cycle in $B_6$. Its fixed locus components are
\begin{equation}
\Phi^\Psi_4\big|_\pm
=-\frac12\beta_\pm^2+y_\pm\Lambda\fluxpole{\pm}\, ,\quad
\Phi^\Psi_2\big|_\pm
=-\frac12y_\pm\beta_\pm\, ,\quad
\Phi^\Psi_0\big|_\pm=-\frac18y_\pm^2\, .
\end{equation}
Since $\Phi^\Psi$ has no six-form component, localization on the six-cycle
gives
\begin{equation}
-\frac12
\left[\beta_+-\pi y_+c_1(\mathcal L)\right]^2
+y_+\Lambda\fluxpole{+}
+\frac12
\left[\beta_--\pi y_-c_1(\mathcal L)\right]^2
-y_-\Lambda\fluxpole{-}
=0\, ,
\end{equation}
in $H^4(B_6,\mathbb R)$ (a similar argument was used in \cite{BenettiGenolini:2023ndb}). Using \eqref{eq:S2-beta-xs} and $y_-=0$, this
reduces immediately to
\begin{equation}\label{eq:S2-cohomological-constraint}
\fluxfib^2+2x\wedge\fluxfib
=2s\,\fluxpole{+}
\quad\text{in }H^4(B_6,\mathbb R)\, .
\end{equation}
For the $S=P=0$ subclass the magnetic flux also satisfies a ``primitivity'' condition on each fixed section. This follows directly from the local supersymmetry equations, as shown in appendix~\ref{app:S2-primitivity}, and in the present notation reads
\begin{equation}\label{eq:S2-primitivity-main}
x\wedge\fluxpole{-}=0\, ,\quad
\left(x+\fluxfib-s\,c_1(\mathcal L)\right)\wedge\fluxpole{+}=0\, .
\end{equation}

There is one further constraint in this subclass. For $S=0$ the electric flux equation in \eqref{ads3beom} must be imposed independently. Integrating it over compact $M_8$ gives
\begin{equation}
\int_{M_8}\me^{6\lambda}F\wedge F=0\, .
\end{equation}
Using the Leray--Hirsch decomposition of the magnetic flux class employed in deriving \eqref{eq:S2-flux-relation}, together with that relation 
and the projective bundle relation
$u^2=u\wedge\pi^*c_1(\mathcal L)$, we find
\begin{align}
0
&=\int_{M_8}
\left[
\frac{\me^{3\lambda}F}{(2\pi\ell_p)^3}
\right]^2 \nn\\
&=\int_{B_6}
\left[
2\,\fluxfib\wedge\fluxpole{-}
+c_1(\mathcal L)\wedge\fluxfib^2
\right]
=\int_{B_6}\fluxfib\wedge
\left(\fluxpole{+}+\fluxpole{-}\right)\, .
\end{align}
Thus, the fluxes obey the compatibility condition
\begin{equation}\label{eq:S2-flux-compatibility-new}
\int_{B_6}\fluxfib\wedge
\left(\fluxpole{+}+\fluxpole{-}\right)=0\, .
\end{equation}
This condition is independent of the non-degeneracy assumptions introduced below.

To solve the remaining constraints, define the map already introduced in the introduction,
\begin{equation}\label{eq:S2-Ln-map}
L_{\fluxfib}:H^2(B_6,\mathbb R)\longrightarrow H^4(B_6,\mathbb R)\, ,\qquad
L_{\fluxfib}(\alpha)=\alpha\wedge\fluxfib\, .
\end{equation}
When $L_{\fluxfib}$ is invertible, \eqref{eq:S2-cohomological-constraint} gives
\begin{equation}\label{eq:S2-x-solution}
x
=s\,L_{\fluxfib}^{-1}(\fluxpole{+})
-\frac12\fluxfib\, .
\end{equation}
Substituting this into the first equation in \eqref{eq:S2-primitivity-main}, and integrating over $B_6$, gives
\begin{equation}\label{eq:S2-s-minus}
s\int_{B_6}L_{\fluxfib}^{-1}(\fluxpole{+})\wedge\fluxpole{-}
=\frac12\int_{B_6}\fluxfib\wedge\fluxpole{-}\, .
\end{equation}
Similarly, substituting \eqref{eq:S2-x-solution} into the second primitivity condition and using the flux relation \eqref{eq:S2-flux-relation} gives
\begin{equation}\label{eq:S2-s-plus}
s\int_{B_6}L_{\fluxfib}^{-1}(\fluxpole{+})\wedge\fluxpole{-}
=-\frac12\int_{B_6}\fluxfib\wedge\fluxpole{+}\, .
\end{equation}
The equality of \eqref{eq:S2-s-minus} and \eqref{eq:S2-s-plus} is precisely the compatibility condition \eqref{eq:S2-flux-compatibility-new}. Thus, provided the common denominator is non-zero,
\begin{equation}\label{eq:S2-s-solution}
s
=
\frac{
\int_{B_6}\fluxfib\wedge\fluxpole{-}}
{
2\int_{B_6}L_{\fluxfib}^{-1}(\fluxpole{+})\wedge\fluxpole{-}}
=-
\frac{
\int_{B_6}\fluxfib\wedge\fluxpole{+}}
{
2\int_{B_6}L_{\fluxfib}^{-1}(\fluxpole{+})\wedge\fluxpole{-}}\, .
\end{equation}

It remains to substitute these relations into the localized central charge. Using \eqref{eq:S2-cohomological-constraint} and the second equation in \eqref{eq:S2-primitivity-main}, the expression \eqref{eq:S2-central-xs} simplifies to
\begin{equation}\label{eq:S2-central-intermediate}
c^{(0)}
=-\int_{B_6}\fluxfib^3
+6s\int_{B_6}\fluxfib\wedge\fluxpole{+}\, .
\end{equation}
Finally, substituting \eqref{eq:S2-s-solution} gives the closed formula
\begin{equation}\label{eq:S2-central-charge-final}
 c^{(0)}
 =
 -\int_{B_6}\fluxfib^3
 +3\,
 \frac{
 \left(
\int_{B_6}\fluxfib\wedge\fluxpole{+}\right)
 \left(
 \int_{B_6}\fluxfib\wedge\fluxpole{-}\right)}
 {
 \int_{B_6}L_{\fluxfib}^{-1}(\fluxpole{+})\wedge\fluxpole{-}}
 \, .
\end{equation}
Thus, the leading central charge is determined entirely by the magnetic flux classes and the cohomology ring of $B_6$.

The closed form \eqref{eq:S2-central-charge-final} applies when $L_{\fluxfib}$ is invertible and
\begin{equation}\label{eq:S2-central-denominator}
\int_{B_6}L_{\fluxfib}^{-1}(\fluxpole{+})\wedge\fluxpole{-}\neq0\, ,
\end{equation}
with the fluxes obeying the compatibility condition \eqref{eq:S2-flux-compatibility-new}. If the denominator in \eqref{eq:S2-central-denominator} vanishes, equations~\eqref{eq:S2-s-minus} and \eqref{eq:S2-s-plus} require both expressions $\int_{B_6}\fluxfib\wedge\fluxpole{\pm}$ to vanish, but do not determine $s$. Likewise, if $L_{\fluxfib}$ is not invertible, the inverse map reduction does not exist. Neither condition by itself implies that the underlying supergravity solution is singular, and this special case instead needs to be analysed separately, as in section~\ref{subsec:MSW}.

\subsection{Gravitational anomaly}\label{subsec:S2-anomaly}
The gravitational anomaly is particularly simple for this class of geometries, and does not require the restriction to $S=P=0$ made in the previous subsection. Substituting the Pontryagin class \eqref{eq:S2-p1-general} into the general formula \eqref{eq:anomaly-cohomological}, and using the fibre-integrated flux class \eqref{eq:S2-fibre-flux-class}, gives
\begin{equation}
\frac{1}{(2\pi\ell_p)^3}
\int_{M_8}\me^{3\lambda}F\wedge p_1(TM_8)
=
\int_{B_6}\fluxfib\wedge
\left[p_1(B_6)+c_1(\mathcal L)^2\right]\, .
\end{equation}
Hence
\begin{equation}\label{eq:S2-anomaly-general}
c_L-c_R
=
-\frac14\int_{B_6}\fluxfib\wedge
\left[p_1(B_6)+c_1(\mathcal L)^2\right]\, .
\end{equation}
Thus, the gravitational anomaly is determined entirely by the fibre magnetic flux and characteristic classes of $B_6$ and the twisting line bundle $\mathcal L$.

\subsection{Calabi--Yau product and the MSW string}\label{subsec:MSW}
An important special case is obtained when the twisting line bundle $\mathcal{L}$ is
trivial and the six-manifold is a Calabi--Yau three-fold $Y_3$, so that
\begin{equation}
M_8=S^2\times Y_3\, .
\end{equation}
This is the familiar near-horizon geometry of the supersymmetric magnetic
string obtained by wrapping an M5-brane on a divisor in $Y_3$. The
corresponding two-dimensional theory is the MSW SCFT introduced in
\cite{MaldacenaStromingerWitten:1997}. We use a left-moving
$\mathcal N=(2,0)$ convention in this paper, whereas the standard MSW
convention is $\mathcal N=(0,4)$ with the supersymmetric sector
right-moving.  Accordingly,
$(c_L,c_R)_{\rm here}=(c_R,c_L)_{\rm MSW}$.
 We also fix the Calabi--Yau complex orientation and denote by
$P\in H^2(Y_3,\mathbb Z)$ the positive divisor class wrapped by the
M5-brane.  With the orientation convention used for the fibre magnetic
flux throughout this paper, the relation between the charges is 
\begin{equation}\label{chargedic}
 \fluxfib=-P\, .
\end{equation}
The $AdS_3\times S^2\times Y_3$ supergravity description, including the
subleading corrections to the central charges, has been studied from a number of complementary viewpoints; see, for example, 
\cite{Kraus_2005,KrausLarsen:2005,CouzensEtAl:2017Ftheory}.

For this solution the magnetic four-form has two legs on the $S^2$ factor.
Consequently its pull-back to either pole section vanishes, and hence
\begin{equation}\label{eq:MSW-pole-fluxes}
\fluxpole{+}=\fluxpole{-}=0\, .
\end{equation}
Equivalently, the magnetic flux class takes the particularly simple form
\begin{equation}
\left[
\frac{\me^{3\lambda}F}{(2\pi\ell_p)^3}
\right]
=
u\wedge\pi^*\fluxfib\, .
\end{equation}
This provides a useful example of the degenerate case discussed at the end
of section~\ref{subsec:S2-central-charge}. In particular, the denominator
in \eqref{eq:S2-central-charge-final} vanishes, so that the inverse map
formula does not apply. Nevertheless, the preceding relation
\eqref{eq:S2-central-intermediate} is perfectly regular and immediately
gives
\begin{equation}\label{eq:MSW-leading-central-charge}
c^{(0)}
=
-\int_{Y_3}\fluxfib^3\, .
\end{equation}
Using the relation \eqref{chargedic}, the fixed complex orientation gives
\begin{equation*}
 c^{(0)}=-\int_{Y_3}(-P)^3=P^3\, ,
\end{equation*}
which is the  cubic intersection term in the MSW central charge.
Thus, the sign in \eqref{eq:MSW-leading-central-charge} is entirely the
charge orientation convention and does not reverse the positive divisor
class $P$ itself.

The gravitational anomaly follows just as directly from
\eqref{eq:S2-anomaly-general}. Since the bundle is trivial and
$c_1(Y_3)=0$, one has
\begin{equation}
p_1(Y_3)=-2c_2(Y_3)\, ,
\end{equation}
and hence, in our flux conventions,
\begin{equation}\label{eq:MSW-gravitational-anomaly}
c_L-c_R
=
\frac12\int_{Y_3}\fluxfib\wedge c_2(Y_3)\, .
\end{equation}
The charge and chirality dictionaries above make the anomaly comparison explicit:
\begin{equation*}
 (c_L-c_R)_{\rm here}
 =-\frac12\,c_2(Y_3)\cdot P
 =(c_R-c_L)_{\rm MSW}\, .
\end{equation*}
Equations~\eqref{eq:MSW-leading-central-charge} and \eqref{eq:MSW-gravitational-anomaly} therefore reproduce the leading cubic term and the difference of the standard MSW central charges, which in the standard MSW chirality convention are
\begin{equation}\label{eq:MSW-standard-central-charges}
 c_L^{\rm MSW}=P^3+c_2(Y_3)\cdot P\, ,\qquad
 c_R^{\rm MSW}=P^3+\frac12c_2(Y_3)\cdot P\, .
\end{equation}
This also fixes the comparison with the five-dimensional Chern--Simons
calculation of \cite{Kraus_2005,KrausLarsen:2005}: in the conventions of
\cite{Kraus_2005} the $SU(2)_R$ symmetry acts on the left-movers, so 
$(c,\widetilde c)$ there correspond to $(c_L,c_R)_{\rm here}$, while their charge
components $q^I$ represent the positive divisor class $P=-\fluxfib$.  The linear Chern--Simons correction then
has precisely the sign in \eqref{eq:MSW-gravitational-anomaly}.
The same M-theory
$AdS_3\times S^2\times Y_3$ geometries, together with their leading and
subleading central charges, were also analysed in
\cite{CouzensEtAl:2017Ftheory}, in the context of M/F-duality for
elliptically fibred Calabi--Yau three-folds.

There is also a direct connection with the localization analysis of
\cite{BenettiGenolini:2024locattr}. Reducing M-theory on $Y_3$ gives
five-dimensional $\mathcal N=2$ ungauged supergravity, whose cubic
couplings are the triple intersection numbers of $Y_3$. Equivariant
localization applied directly to the $AdS_3\times S^2$ black string
near-horizon geometry in \cite{BenettiGenolini:2024locattr} reproduces the
usual attractor equations and the cubic leading central charge.\footnote{%
In section~4 of \cite{BenettiGenolini:2024locattr} the possibility of an
additional solution with non-zero parameters $x^I$ was left open when the
matrix $C_{IJK}\mathfrak p^K$ is degenerate.  For the Calabi--Yau MSW
solution it is the positive divisor class $P$, not our magnetic class
$\fluxfib=-P$, that lies in the K\"ahler cone.  Wedging with
$P$ is the Lefschetz map
$H^{1,1}(Y_3)\rightarrow H^{2,2}(Y_3)$ and is an isomorphism by the Hard
Lefschetz theorem.  Replacing $P$ by $-P$ only multiplies this map by an
overall minus sign and therefore does not affect its invertibility.  Thus,
the relevant matrix is non-degenerate in either charge convention, fixing
$x^I=0$ for this solution.}
The present computation may thus be viewed as an eleven-dimensional
counterpart of that localization calculation, i.e., rather than first reducing
on the Calabi--Yau three-fold, the equivariant localization is performed
directly on the full internal space $S^2\times Y_3$.

\section{Specializations and explicit $S^2$ bundle families}\label{sec:S2specializations}
In this section, we apply the general results of section~\ref{sec:S2B6} to more concrete choices of the six-dimensional base and magnetic fluxes. We begin with
$B_6=T^2\times KE_4$, a topology motivated by the explicit compact solutions of
\cite{Gauntlett:2006qw}, but initially impose only the cohomological and flux constraints derived in the previous section. The resulting central charge therefore applies to a considerably more general set of flux data than the presently known analytic solutions. This is analogous to the $AdS_5\times M_6$ analysis of \cite{BenettiGenolini:2023ndb}, where equivariant localization on a fixed $S^2$ bundle topology gives observables for general compatible fluxes, without requiring an explicit supergravity solution. In the following subsection we specialize these results to the known analytic family, while we later consider a numerical solution outside that family with fully quantized fluxes.

\subsection{General fluxes on $T^2\times KE_4$}\label{subsec:T2KE-general}
Let
\begin{equation}
B_6=T^2\times KE_4\, ,
\end{equation}
where $KE_4$ is a compact K\"ahler--Einstein four-manifold, and set
$r=b_2(KE_4)$ as the second Betti number. Throughout this subsection, we assume
$ b_1(KE_4)=0$, which is necessarily the case for the positively curved K\"ahler--Einstein surfaces used in the explicit family of
the next subsection. In particular, this implies that there are no odd-degree cohomology class contributions to the flux map below (by the K\"unneth formula). 
 We take the twisting line bundle $\mathcal L$ to be pulled back from $KE_4$, as in the geometries of interest, but for now do not impose the additional relations between the twisting and magnetic fluxes that hold for the known explicit supergravity solutions.

Let $\tau\in H^2(T^2,\mathbb Z)$ and
$\Omega_4\in H^4(KE_4,\mathbb Z)$ be normalized so that
\begin{equation}
\int_{T^2}\tau=1\, ,\qquad
\int_{KE_4}\Omega_4=1\, ,
\end{equation}
and choose an integral basis
$\Psi^a\in H^2(KE_4,\mathbb Z)$, $a=1,\ldots,r$. We denote the intersection form by
\begin{equation}\label{eq:T2KE-intersection}
I^{ab}
\equiv
\int_{KE_4}\Psi^a\wedge\Psi^b\, ,
\end{equation}
so that in cohomology
\begin{equation}
\Psi^a\wedge\Psi^b=I^{ab}\mskip1mu\Omega_4\, .
\end{equation}
For  $U=(U_a)$ and $V=(V_a)$ representing degree two cohomology classes, we write
\begin{equation}
U\cdot V\equiv I^{ab}U_aV_b\, .
\end{equation}

With this notation to hand, we next expand the fibre flux, the flux on the minus section and the twisting class as
\begin{align}\label{eq:T2KE-flux-expansion}
\fluxfib
&=
N_0\tau+N_a\Psi^a\, , \quad
\fluxpole{-}
=
M_0\Omega_4+M_a\Psi^a\wedge\tau\, ,\quad
c_1(\mathcal L)
=
C_a\Psi^a\, .
\end{align}
In the next subsection, we will take $N_a$, $M_a$ and $C_a$ to be mutually proportional in order to recover results for explicit families of K\"ahler--Einstein supergravity solutions. Within the $b_1(KE_4)=0$ cohomology sector just specified, no further restriction is imposed here on the magnetic fluxes. 
The topological flux relation \eqref{eq:S2-flux-relation} determines the flux on the other pole section:
\begin{equation}\label{eq:T2KE-plus-flux}
\fluxpole{+}
=
\left(M_0+C\cdot N\right)\Omega_4
+
\left(M_a+N_0C_a\right)\Psi^a\wedge\tau\, .
\end{equation}
The compatibility condition \eqref{eq:S2-flux-compatibility-new} then reduces to
\begin{equation}\label{eq:T2KE-compatibility}
N_0M_0+N\cdot M+N_0C\cdot N=0\, .
\end{equation}

We next evaluate the inverse map entering the central charge
\eqref{eq:S2-central-charge-final}. Because $b_1(KE_4)=0$ removes the mixed K\"unneth components, the map written below is the full cohomology map
$L_{\fluxfib}:H^2(B_6)\to H^4(B_6)$ for the product topology under consideration. We consider the non-degenerate branch
\begin{equation}
N_0\neq0\, ,\qquad N^2\equiv N\cdot N\neq0\, ,
\end{equation}
which includes the explicit solutions discussed in section~\ref{subsec:T2KE-explicit}, and write
\begin{equation}
L_{\fluxfib}^{-1}(\fluxpole{+})
=
A_0\tau+A_a\Psi^a\, .
\end{equation}
The equation
\begin{equation}
\left(A_0\tau+A_a\Psi^a\right)\wedge\fluxfib
=
\fluxpole{+}
\end{equation}
has components
\begin{equation}
A\cdot N=M_0+C\cdot N\, ,\qquad
N_0A_a+A_0N_a=M_a+N_0C_a\, .
\end{equation}
Solving gives
\begin{align}\label{eq:T2KE-A-solution}
A_0
&=
\frac{M\cdot N-N_0M_0}{N^2}\, , \qquad 
A_a
=
C_a+\frac{M_a}{N_0}
-\frac{N_a}{N_0N^2}
\left(M\cdot N-N_0M_0\right)\, .
\end{align}
It follows that
\begin{equation}\label{eq:T2KE-denominator}
\int_{B_6}
L_{\fluxfib}^{-1}(\fluxpole{+})
\wedge\fluxpole{-}
=
\frac{\Delta}{N_0N^2}\, ,
\end{equation}
where
\begin{equation}\label{eq:T2KE-Delta}
\Delta
\equiv
N_0N^2\,C\cdot M
+M^2N^2
-\left(N\cdot M-N_0M_0\right)^2\, ,
\qquad
M^2\equiv M\cdot M\, .
\end{equation}

The remaining quantities appearing in
\eqref{eq:S2-central-charge-final} are particularly simple. We have
\begin{equation}
\int_{B_6}\fluxfib^3
=
3N_0N^2\, ,
\end{equation}
while using \eqref{eq:T2KE-compatibility} gives
\begin{equation}
\int_{B_6}\fluxfib\wedge\fluxpole{-}
=
-N_0C\cdot N\, ,\qquad
\int_{B_6}\fluxfib\wedge\fluxpole{+}
=
N_0C\cdot N\, .
\end{equation}
Substitution into the general formula
\eqref{eq:S2-central-charge-final} therefore gives
\begin{equation}\label{eq:KE4-central-oriented}
c^{(0)}
=
-3N_0N^2
-
\frac{3N_0^3N^2(C\cdot N)^2}{\Delta}
\, .
\end{equation}
This expression is valid when the compatibility condition
\eqref{eq:T2KE-compatibility} holds and $\Delta\neq0$, on the non-degenerate branch considered above. As in section~\ref{subsec:S2-central-charge}, failure of these assumptions does not imply that the underlying supergravity solution is singular, i.e., rather, the corresponding degenerate cases must be treated using the more general relations that appear before the inverse map formula.

Equation~\eqref{eq:KE4-central-oriented} is a cohomological result for general compatible flux data on $T^2\times KE_4$, subject to $b_1(KE_4)=0$ and the stated non-degeneracy assumptions. In particular, no relation has been imposed between the flux vector $N_a$ and the canonical class of $KE_4$. The known analytic solutions form a special locus in this larger space of flux data, to which we turn in the next subsection.

For completeness, the gravitational anomaly also follows immediately from
\eqref{eq:S2-anomaly-general}. Since both $p_1(B_6)=p_1(KE_4)$ and
$c_1(\mathcal L)^2$ are pulled back from $KE_4$, only the $N_0\tau$
component of $\fluxfib$ contributes, giving
\begin{equation}\label{eq:T2KE-anomaly-general}
c_L-c_R
=
-\frac{N_0}{4}
\int_{KE_4}
\left[
p_1(KE_4)+c_1(\mathcal L)^2
\right]\, .
\end{equation}

\subsection{The known analytic $T^2\times KE_4$ solutions}
\label{subsec:T2KE-explicit}
We now specialize the general result of section~\ref{subsec:T2KE-general}
to the explicit compact solutions of \cite{Gauntlett:2006qw}. For
$B_6=T^2\times KE_4$ with $KE_4$ positively curved these eleven-dimensional backgrounds may be reduced
on one circle of the $T^2$ and T-dualized on the other to give the Type IIB
solutions constructed and analysed globally in \cite{Gauntlett:2006af}.
The explicit ansatz imposes considerably stronger restrictions on the
cohomology classes than those assumed in the previous subsection.

Indeed, the dependence of the local solution on $KE_4$ is entirely through
its K\"ahler form. Since the K\"ahler--Einstein condition relates the K\"ahler form to the  Ricci form, with the latter representing $2\pi c_1(KE_4)$ in cohomology, all
$H^2(KE_4)$ components of the magnetic flux are therefore proportional to
the first Chern class. To make this explicit, let $m$ denote the Fano index
of $KE_4$ and introduce the primitive integral class $\nu$ by
\begin{equation}\label{eq:T2KE-nu-def}
c_1(KE_4)=m\mskip1mu\nu \in H^2(KE_4,\Z)\, ,\qquad
\nu=\nu_a\Psi^a\, .
\end{equation}
We also define
\begin{equation}\label{eq:T2KE-S-def}
S\equiv\int_{KE_4}\nu^2
=\nu\cdot\nu
=\frac{M_{\mathrm{KE}}}{m^2}\, ,\qquad
M_{\mathrm{KE}}\equiv\int_{KE_4}c_1(KE_4)^2\, .
\end{equation}
For the $S^2$ bundle appearing in the explicit solution, the twisting line
bundle $\mathcal{L}$ is the canonical bundle, and hence in the conventions of
section~\ref{sec:S2B6}
\begin{equation}\label{eq:T2KE-explicit-twist}
c_1(\mathcal L)=-c_1(KE_4)=-m\nu\, .
\end{equation}
Thus, the vector $C_a$ introduced in \eqref{eq:T2KE-flux-expansion} is
\begin{equation}
C_a=-m\nu_a\, .
\end{equation}

The same structure of the flux ansatz implies that the $KE_4$ components
of both magnetic flux classes in \eqref{eq:T2KE-flux-expansion} are also
proportional to $\nu$. The global analysis of
\cite{Gauntlett:2006qw,Gauntlett:2006af}  parameterizes the regular solutions by positive integers $p,q$, which we take to be coprime when imposing flux quantization, together with the overall flux integer $n$. To compare to the above references, we define
\begin{equation}\label{eq:T2KE-kappa-def}
\kappa\equiv\frac{n}{h}\, ,\quad
h\equiv\gcd(S,q)\, .
\end{equation}
In the projective bundle orientation used in this paper, the corresponding
fluxes are
\begin{equation}\label{eq:KE4-signed-dictionary}
N_0=-q\, ,\quad
N_a=-\kappa q\,\nu_a\, ,\quad
M_0=\kappa p\mskip1mu S\, ,\quad
M_a=-(p+mq)\nu_a\, .
\end{equation}
Equivalently, the three cohomology classes entering the general analysis
are then
\begin{align}\label{eq:T2KE-explicit-classes}
\fluxfib
&=-q\,\tau-\kappa q\,\nu\, ,\nonumber\\
\fluxpole{-}
&=\kappa p\mskip1mu S\,\Omega_4-(p+mq)\nu\wedge\tau\, ,\\
c_1(\mathcal L)
&=-m\mskip1mu \nu\, .\nonumber
\end{align}
The minus signs in these formulae arise from our choice of orientation for
the $S^2$ bundle and magnetic flux periods.

It is useful to check directly how the topology relates the fluxes at the
two poles. Substituting \eqref{eq:T2KE-explicit-classes} into
\eqref{eq:S2-flux-relation} gives
\begin{equation}\label{eq:T2KE-explicit-plus}
\fluxpole{+}
=
\kappa(p+mq)S\,\Omega_4
-p\,\nu\wedge\tau\, .
\end{equation}
Thus, the two pole sections naturally involve the combinations $p$ and
$p+mq$ that appear in the global description of the solution. Notice also
that the compatibility condition \eqref{eq:T2KE-compatibility} is
automatically satisfied:
\begin{equation}
-\kappa pq\mskip1mu S
+\kappa q(p+mq)S
-m\kappa q^2S
=0\, .
\end{equation}
The explicit supergravity ansatz therefore selects a special
one-dimensional direction in each of the $H^2(KE_4)$ flux vectors appearing
in the more general analysis of section~\ref{subsec:T2KE-general}. This is
closely analogous to the $AdS_5\times M_6$ examples of
\cite{BenettiGenolini:2023ndb}, where the known explicit solutions likewise
restrict otherwise general magnetic fluxes to the direction determined by
the canonical class.

We can now evaluate the general central charge
\eqref{eq:KE4-central-oriented}. From
\eqref{eq:KE4-signed-dictionary} and \eqref{eq:T2KE-S-def} one finds
\begin{equation}
N^2=\kappa^2q^2S\, ,\quad
C\cdot N=m\kappa qS\, ,\quad
M^2=(p+mq)^2S\, ,
\end{equation}
together with
\begin{equation}
N\cdot M-N_0M_0
=
\kappa q(2p+mq)S\, .
\end{equation}
The quantity \eqref{eq:T2KE-Delta} therefore becomes
\begin{equation}\label{eq:T2KE-explicit-Delta}
\Delta
=
-\kappa^2q^2S^2
\left(
3p^2+3mpq+m^2q^2
\right)\, .
\end{equation}
Substitution into \eqref{eq:KE4-central-oriented} and using $\kappa=n/h$ and
$S=M_{\mathrm{KE}}/m^2$ then gives
\begin{equation}\label{eq:T2KE-IIB-central}
c^{(0)}
=
\frac{9pq^2(p+mq)}
{3p^2+3mpq+m^2q^2}
\frac{M_{\mathrm{KE}}q}{m^2h^2}\,n^2
\, .
\end{equation}
This precisely reproduces the holographic central charge first obtained
from the Type IIB dual geometry in \cite{Gauntlett:2006af}. The agreement
is a non-trivial check of the localization formula: the rather intricate
rational expression \eqref{eq:T2KE-IIB-central} follows here entirely from
the topology and quantized magnetic fluxes, without using the explicit
metric functions of the supergravity solution.

The gravitational anomaly is similarly immediate. Since
$c_1(\mathcal L)^2=c_1(KE_4)^2$ and $N_0=-q$,
\eqref{eq:T2KE-anomaly-general} gives
\begin{equation}\label{eq:KE4-anomaly}
c_L-c_R
=
\frac{q}{4}
\left[
P_1(KE_4)+M_{\mathrm{KE}}
\right]\, ,
\qquad
P_1(KE_4)\equiv\int_{KE_4}p_1(KE_4)\, .
\end{equation}
Equivalently, using
$P_1(KE_4)=3\sigma(KE_4)$ and
$M_{\mathrm{KE}}=2\chi(KE_4)+3\sigma(KE_4)$,
\begin{equation}
c_L-c_R
=
\frac{q}{2}
\left[
3\sigma(KE_4)+\chi(KE_4)
\right]\, .
\end{equation}

\subsection{A flux-quantized non-polynomial $AdS_3$ solution}
\label{subsec:loc-np}
We next consider a numerical solution of the same general $S^2$ bundle
system, but outside the known polynomial family. The base is
$B_6=S^2_1\times S^2_2\times T^2$, so this is a special case of
section~\ref{subsec:T2KE-general}, with $KE_4=S^2_1\times S^2_2$. Unlike the
explicit analytic family of section~\ref{subsec:T2KE-explicit}, however, the
quantized magnetic flux is not confined to the canonical-class direction.
The example therefore provides a concrete realization of the more general
cohomological flux sector.

We use the cohomogeneity one system of \cite{Gauntlett:2006qw}, but do not
impose the polynomial ansatz used there to construct explicit compact
solutions. Here we focus on a single compact numerical solution, prove below
that its quantized flux sector admits no regular compact polynomial
representative of this master equation, and use it to illustrate the general localization formulae in this more
general flux sector.
A more systematic analysis of the unrestricted equation, including analytic
existence beyond the polynomial sector, the curvature-sign classification
and further flux-quantized families, will be presented in \cite{Gregory:2026nonpolynomial}.

The cohomogeneity one ansatz for the metric is \cite{Gauntlett:2006qw}
\begin{equation}\label{eq:loc-np-metric-ansatz}
\dd s^2_{11}
=
\me^{2\lambda}
\Big[
\dd s^2(AdS_3)
+\sum_{i=1}^3 h_i(r)\,\dd s^2(C_i)
+f_3(r)\,\dd r^2
+f_4(r)\,(D\psi)^2
\Big]\, ,
\end{equation}
with
\begin{equation}\label{eq:loc-np-base}
(C_1,C_2,C_3)=(S^2_1,S^2_2,T^2),
\qquad
(k_1,k_2,k_3)=(1,1,0),
\end{equation}
where the $C_i$ are constant curvature K\"ahler surfaces with K\"ahler forms
$J_i$, and the circle coordinate has period $\psi\sim\psi+2\pi$. The connection obeys
\begin{equation}
\dd(D\psi)=J_1+J_2\, .
\end{equation}
The four-form is expanded in the corresponding invariant basis constructed
from $J_i$, $\dd r$ and $D\psi$. Supersymmetry reduces the radial dependence
of both the metric and flux to a single function $H(r)$, together with
constants $c_i,d_i$. We keep the same geometric ansatz and compact endpoint conditions, but do not require $H$ to be a polynomial. Once $H$ is known,
the remaining metric functions are reconstructed algebraically.

Details of the numerical construction and flux normalization are collected in
Appendix~\ref{app:nonpolynomial-details}. We use the radial translation
freedom to set $c_3=0$ throughout. It is useful
to define
\begin{equation}\label{eq:loc-np-Q-definitions}
Q_1=-4(d_1+c_1r)\,,\quad
Q_2=-4(d_2+c_2r)\,,\quad
Q_3=r^2-4d_3\,,\quad
P=Q_1Q_2Q_3\,.
\end{equation}
The unrestricted supersymmetry system then reduces to the single master
equation
\begin{equation}\label{eq:loc-np-master}
-(H')^2+2HH''-(Q_1+Q_2)H+\frac14P=0\,.
\end{equation}
A compact solution is defined on an interval $[r_-,r_+]$ with
\begin{equation}\label{eq:loc-np-boundary}
H(r_\pm)=0\,,\qquad H(r)>0\quad(r_-<r<r_+)\,.
\end{equation}
At a simple endpoint, the equation itself fixes the slope,
\begin{equation}\label{eq:loc-np-endpoint-slope}
\bigl[H'(r_\pm)\bigr]^2=\frac14P(r_\pm)\,,
\end{equation}
with the positive sign at $r_-$ and the negative sign at $r_+$ for a positive
solution between the two roots. Together with $\psi\sim\psi+2\pi$, this
endpoint condition guarantees that the two-dimensional
fibre collapses smoothly. Regularity of the solution also requires
the sign conditions in the interior of the interval that we state below.

While the parameters and endpoint values of the compact numerical solution, as well as the endpoint derivatives and remaining eleven-dimensional checks, are given in Appendix~\ref{app:nonpolynomial-details}, here we present only the data relevant for the localization analysis.
Introducing
\begin{equation}\label{eq:loc-np-D}
D=2H''-(Q_1+Q_2)\,,
\end{equation}
one has
\begin{equation}\label{eq:loc-np-h}
h_i=-\frac{D}{4Q_i}\,,\quad i=1,2,3\,,
\end{equation}
and
\begin{equation}\label{eq:loc-np-f}
f_3=-\frac{D}{16H}\,,\quad
f_4=-\frac{HD}{P}\,,\quad
\ex^{6\lambda}=\frac{4P}{D^2}\,.
\end{equation}
For the numerical solution,
\begin{equation}\label{eq:loc-np-signs}
H>0\quad(r_-<r<r_+)\,,\quad
Q_i>0\,,\quad D<0\quad(r_-\leq r\leq r_+)\,.
\end{equation}
Hence every metric coefficient is positive in the interior,
and the simple endpoint zeros close the two-dimensional fibre smoothly at
both ends.

As a check on the numerical solution, we also integrated inward independently from the two regular endpoints, using the corresponding local expansions, and matched $H$ and $H'$ at an interior point. Varying the endpoint cut-offs and numerical precision leaves the endpoint data and flux periods unchanged at the quoted precision, while $H$, $Q_i$ and $-D$ remain positive throughout the interval. Substitution of the reconstructed metric and four-form into the eleven-dimensional equations shows that the Bianchi identity, four-form equation and Einstein equations are satisfied to approximately $10^{-15}$, $10^{-9}$ and $10^{-10}$, respectively. After fixing the overall scale and torus area, the remaining four flux periods agree with integers to approximately $3\times10^{-13}$.

The flux data for this solution are what makes it particularly interesting. Let
$\omega_1,\omega_2,\tau$ be unit-period integral generators of the two
sphere factors and the torus. The twisting line bundle $\mathcal{L}$ is
\begin{equation}\label{eq:loc-np-twist}
c_1(\mathcal L)=2(\omega_1+\omega_2)=c_1(B_6)\, ,
\end{equation}
so here $\mathcal L$ is anti-canonical. Recall that $\mathbb P(\mathcal O\oplus\mathcal L)$ and $\mathbb P(\mathcal O\oplus\mathcal L^{-1})$ describe the same $S^2$ bundle, with the two fixed sections exchanged. Thus, the opposite choice of $\mathcal L$ compared with section~\ref{subsec:T2KE-explicit} is a convention for labelling the two poles. In the notation of
section~\ref{sec:S2B6}, write
\begin{align}\label{eq:loc-np-flux-classes}
\fluxfib
&=N_1\omega_1+N_2\omega_2+N_0\tau\,,\nonumber\\
\fluxpole{-}
&=M_0\,\omega_1\wedge\omega_2
 +M_1\,\omega_1\wedge\tau
 +M_2\,\omega_2\wedge\tau\, .
\end{align}
Thus, $N_0,N_1,N_2$ are the three fibre-integrated periods, while
$M_0,M_1,M_2$ are the periods on one pole section. Restoring the overall
scale and fixing the torus area as described in
Appendix~\ref{app:nonpolynomial-details} gives
\begin{align}
(N_0,N_1,N_2 ; M_0,M_1,M_2)
=
(-3,-10,-11 ; -28,10,10)\,.
\end{align}
Thus,
\begin{align}
\fluxfib=-10\omega_1-11\omega_2-3\tau \, .
\end{align}
In particular, the $S^2_1\times S^2_2$ component is not proportional to
$\omega_1+\omega_2$. This is precisely the more general flux direction
allowed by section~\ref{subsec:T2KE-general} but absent from the analytic
family of section~\ref{subsec:T2KE-explicit}. The integer vector also satisfies the compatibility condition
\eqref{eq:T2KE-compatibility}. Recall that the flux on the opposite fixed section is not
independent, i.e., using
$
\fluxpole{+}-\fluxpole{-}=c_1(\mathcal L)\wedge\fluxfib$,
one finds, in the same ordering of four-cycles,
\begin{equation}\label{eq:loc-np-north}
(M_0^+,M_1^+,M_2^+)=(-70,4,4).
\end{equation}
These opposite-pole periods follow from the projective bundle relation
and verify that the flux conventions are consistent -- they do not provide an additional
test of the interior solution.

The central charge may be expressed using either the
Brown--Henneaux endpoint formula or the localized flux formula, giving
\begin{align}
c_{\mathrm{sol}}^{(0)}
\simeq1523.3424371\, ,\qquad
c_{\mathrm{loc}}^{(0)}
=\frac{26215200}{17209}
\simeq1523.3424371\, .
\end{align}
The equality is an algebraic consequence of the endpoint compatibility
conditions, as proved in \cite{Gregory:2026nonpolynomial}. Therefore, it
holds independently of whether those endpoint data admit a regular
interior solution. This example illustrates the flux formula derived in
section~\ref{subsec:S2-central-charge}, while existence and regularity require
the separate boundary-value analysis. The gravitational anomaly follows
independently from the topological formula,
\begin{align}
c_L-c_R=-2N_0=6\,.
\end{align}
\subsubsection*{Genuine non-polynomiality}
\label{subsec:loc-np-polynomial}
Non-polynomiality follows directly from the integral flux data. For a regular
compact polynomial solution of \eqref{eq:loc-np-master} with the endpoint and
positivity conditions \eqref{eq:loc-np-boundary} and
\eqref{eq:loc-np-signs}, we show in
Appendix~\ref{app:polynomial-elimination} that
\begin{equation}\label{eq:loc-np-poly-equal-flux}
M_1=M_2\quad\Longrightarrow\quad N_1=N_2\,.
\end{equation}
The proof uses the polynomial coefficient equations and regular endpoint
slopes. Every polynomial solution has degree at most three, and the argument
includes all lower degree cases without assuming a generic cubic branch.
The solution has
\begin{equation}\label{eq:loc-np-poly-benchmark}
M_1=M_2=10\,,\qquad N_1=-10\neq-11=N_2\,,
\end{equation}
so its quantized fluxes admit no regular compact polynomial
representative within this ansatz. 

\section{Discussion}\label{sec:discussion}

In this paper, we have developed equivariant localization for a general
class of supersymmetric $AdS_3\times M_8$ backgrounds of
eleven-dimensional supergravity. Subject to the global exactness condition
\eqref{eq:global-potential-condition}, the supersymmetry equations naturally
organize the fluxes, calibrations and warped volume into equivariantly
closed polyforms for the R-symmetry action. Their integrals may therefore
be evaluated in terms of fixed point data, giving expressions for
holographic observables without requiring an explicit solution of the
full supergravity equations.

For the $S^4$ bundle solutions associated with M5-branes wrapped on
coassociative four-cycles, localization reproduces the leading central
charge, including its topological coefficient, directly from the fixed
point geometry. The gravitational anomaly is similarly expressed in terms
of characteristic numbers of the four-dimensional base. For the more
general $S^2$ bundles over $B_6$, the relevant data are the fibre flux
$\fluxfib$, the two pole fluxes $\fluxpole{\pm}$, the twisting line bundle
$\mathcal L$, and the intersection structure of $B_6$. In the $S=P=0$
subclass, and on the non-degenerate branch discussed in
section~\ref{sec:S2B6}, the leading central charge is determined entirely
by these cohomological data through the map
\begin{equation}
L_{\fluxfib}:H^2(B_6,\mathbb R)\longrightarrow H^4(B_6,\mathbb R)\, ,
\qquad
L_{\fluxfib}(\alpha)=\alpha\wedge\fluxfib\, .
\end{equation}
The gravitational anomaly takes the particularly simple form
\begin{equation}
c_L-c_R
=
-\frac14\int_{B_6}\fluxfib\wedge
\left[p_1(B_6)+c_1(\mathcal L)^2\right]\, .
\end{equation}

The specializations in section~\ref{sec:S2specializations} illustrate these
general results in two rather different settings. For
$B_6=T^2\times KE_4$, the cohomological analysis allows general compatible
magnetic fluxes, while the known analytic solutions restrict the
$H^2(KE_4)$ components to the canonical class direction. On this locus, the
localization formula reproduces precisely the central charge previously
obtained from the Type IIB dual solutions. The numerical
$S^2\times S^2\times T^2$ solution instead realizes flux outside this
restricted locus. For this ansatz the Brown--Henneaux expression reduces to
endpoint data, and its equality with the localized flux formula follows
algebraically from the endpoint compatibility conditions. The existence and
regularity of the corresponding interior solution are determined
separately by the numerical boundary value problem. Moreover,
Appendix~\ref{app:polynomial-elimination} shows that any regular compact
polynomial solution in this ansatz with $M_1=M_2$ must also have
$N_1=N_2$. The flux vector of the numerical solution violates this
condition, and hence admits no polynomial representative within the ansatz.

Localization by itself does not determine whether a given set of
topological and flux data is realized by a smooth supersymmetric
background. Existence and regularity remain separate questions, as the
non-polynomial example makes clear. A natural problem is
therefore to understand more systematically which flux sectors admitted
by the localization constraints are realized by supergravity solutions. It would be interesting to
understand the corresponding existence problem more generally, and in
particular how much information about the space of supersymmetric $AdS_3$
solutions can be extracted directly from the equivariant and cohomological
data.

\section*{Acknowledgements}

\noindent 
The work of JFS was supported in part by STFC grant  ST/X000761/1. 
CCN is supported by the Martingale Postgraduate Foundation. 

\appendix

\section{A constraint on the magnetic flux}\label{app:S2-primitivity}

In this appendix, we derive the  condition on the magnetic flux that is used in section~\ref{sec:S2B6}. In the conventions used throughout the paper, the $AdS_3$ radius is set to one, so the constant $m$ appearing in the supersymmetry equations of \cite{Ashmore:2022ydf} is fixed to $m=-\frac12$. We restrict to the $S=P=0$ subclass, where the $SU(3)$ structure forms of \cite{Ashmore:2022ydf} satisfy
\begin{equation}\label{eq:app-S2-SU3-forms}
\Psi=-\frac12j^2-\sin\alpha\,j\wedge e^{78}\, ,\qquad
*_8\Psi=-j\wedge e^{78}-\frac12\sin\alpha\,j^2\, ,
\end{equation}
and
\begin{equation}\label{eq:app-S2-phi-form}
\phi=-\cos\alpha\,j\wedge e^7\, .
\end{equation}
Here $j$ is the real two-form defining the $SU(3)$ structure on the six-dimensional space orthogonal to $e^7,e^8$, and $e^{78}\equiv e^7\wedge e^8$. 
The supersymmetry conditions include the differential relation
\begin{equation}\label{eq:app-S2-differential-condition}
\me^{-6\lambda}\diff\left(\me^{6\lambda}\phi\right)
=-4m\Psi+\sin\alpha\,F-*_8F\, ,
\end{equation}
and the algebraic relation
\begin{equation}\label{eq:app-S2-algebraic-condition}
2m=\frac13K\hook f-\frac16(*_8\Psi)\hook F\, .
\end{equation}
These are the corresponding conditions in equation~(4.25) of \cite{Ashmore:2022ydf}, written in our conventions.

At either fixed section $B_6^\pm$ we decompose the magnetic four-form as
\begin{equation}\label{eq:app-S2-F-decomposition}
F=F_4+e^{78}\wedge F_2\, ,
\end{equation}
where $F_4$ and $F_2$ have legs only along the six-dimensional distribution. In particular,
\begin{equation}
F\big|_{B_6^\pm}=F_4\, .
\end{equation}
Choosing the orientation so that $\vol_8=\vol_6\wedge e^{78}$, the Hodge star decomposes as
\begin{equation}\label{eq:app-S2-Hodge-decomposition}
*_8F_4=(*_6F_4)\wedge e^{78}\, ,\qquad
*_8(e^{78}\wedge F_2)=*_6F_2\, .
\end{equation}

On $B_6^\pm$ one has $\cos\alpha=0$, so the pullback of \eqref{eq:app-S2-phi-form} vanishes. Pull-back commutes with the exterior derivative, and hence the left-hand side of \eqref{eq:app-S2-differential-condition} also vanishes on the fixed section. Using
\begin{equation}
\Psi\big|_{B_6^\pm}=-\frac12j^2\, ,
\end{equation}
together with \eqref{eq:app-S2-Hodge-decomposition}, the pullback of \eqref{eq:app-S2-differential-condition} gives
\begin{equation}
0=2m j^2+\sigma F_4-*_6F_2\, .
\end{equation}
Here we have  introduced
\begin{equation}
\sigma\equiv\sin\alpha\big|_{B_6^\pm}=\pm1\, .
\end{equation}
Since $*_6j^2=2j$, this is equivalent to
\begin{equation}\label{eq:app-S2-F2}
F_2=4m j+\sigma *_6F_4\, .
\end{equation}

We now use the algebraic supersymmetry condition \eqref{eq:app-S2-algebraic-condition}. On the fixed set $K=0$, while \eqref{eq:app-S2-SU3-forms} and \eqref{eq:app-S2-F-decomposition} give
\begin{equation}\label{eq:app-S2-PsiF-contraction}
(*_8\Psi)\hook F
=-j\hook F_2-\frac12\sigma j^2\hook F_4\, .
\end{equation}
It follows that
\begin{equation}\label{eq:app-S2-algebraic-expanded}
12m=j\hook F_2+\frac12\sigma j^2\hook F_4\, .
\end{equation}
Substituting \eqref{eq:app-S2-F2} and using $j\hook j=3$ gives
\begin{equation}
12m=12m+\sigma\Big[j\hook(*_6F_4)+\frac12j^2\hook F_4\Big]\, ,
\end{equation}
and hence
\begin{equation}\label{eq:app-S2-primitivity-contraction}
j\hook(*_6F_4)+\frac12j^2\hook F_4=0\, .
\end{equation}

To interpret this relation, we use the standard pointwise inner product on forms,
\begin{equation}
\alpha\wedge *_6\beta=\langle\alpha,\beta\rangle\mskip1mu\vol_6\, ,\qquad
\alpha\hook\beta=\langle\alpha,\beta\rangle\, ,
\end{equation}
together with
\begin{equation}
*_6j=\frac12j^2\, .
\end{equation}
These imply
\begin{equation}\label{eq:app-S2-Hodge-identity}
j\hook(*_6F_4)
=\langle j,*_6F_4\rangle
=\langle *_6j,F_4\rangle
=\frac12j^2\hook F_4\, .
\end{equation}
Equation~\eqref{eq:app-S2-primitivity-contraction} therefore implies
\begin{equation}
j^2\hook F_4=0\, ,\qquad
j\hook(*_6F_4)=0\, .
\end{equation}
The second condition is equivalently
\begin{equation}\label{eq:app-S2-primitivity}
j\wedge F_4=0\, ,
\end{equation}
since
\begin{equation}
j\wedge F_4
=\langle j,*_6F_4\rangle\vol_6\, .
\end{equation}
Thus, the magnetic four-form restricted to either fixed section satisfies the constraint $j\wedge F_4=0$  with respect to the corresponding $SU(3)$ structure two-form.

Finally, on the fixed sections the parameterization of \cite{Ashmore:2022ydf} gives
\begin{equation}
\omega\big|_{B_6^\pm}=\sigma j\, .
\end{equation}
Moreover, the differential condition for $\omega$ implies that
$\me^{3\lambda}\omega$ is closed after restriction to the fixed sections. We may therefore define
\begin{equation}
\beta_\pm\equiv
\Big[\me^{3\lambda}\omega\big|_{B_6^\pm}\Big]
\in H^2(B_6,\mathbb R)\, .
\end{equation}
Using the flux classes $\fluxpole{\pm}$ defined in
\eqref{eq:S2-pole-flux-classes}, the primitivity condition
\eqref{eq:app-S2-primitivity} becomes
\begin{equation}\label{eq:app-S2-primitivity-final}
\beta_\pm\wedge\fluxpole{\pm}=0\, .
\end{equation}
Note that this condition follows directly from the local supersymmetry equations in the $S=P=0$ subclass, i.e., no localization argument is required.

\section{Numerical solution}
\label{app:nonpolynomial-details}

This appendix gives further details of the numerical solution discussed in section~\ref{subsec:loc-np}. We record the endpoint data and flux normalization, together with the direct Brown--Henneaux calculation and the argument excluding polynomial solutions. The metric ansatz, master equation and boundary conditions are given in the main text.

\subsection{Numerical solution and regularity}
\label{subsec:loc-np-numerics}

Solving the two-point boundary value problem numerically gives
\begin{align}
c_1=c_2&=1.02\, ,
&
c_3&=0\, ,
\notag\\
d_1&=-1\, ,
&
d_2&=-0.98489221891510\, ,
&
d_3&=0.13201704744117\, ,
\label{eq:loc-np-parameters}
\end{align}
\phantomsection\label{eq:numerical-parameters}
with endpoints
\begin{equation}
r_-=0.76580730923010\, ,
\quad
r_+=0.85624260578730\, .
\label{eq:loc-np-endpoints}
\end{equation}
\phantomsection\label{eq:numerical-endpoints}
At the precision quoted here, the endpoint derivatives are
\begin{align}
H'(r_-)&\simeq0.1020651521\, ,
&
H'(r_+)&\simeq-0.1076349621\, ,
\notag\\
H''(r_-)&\simeq-2.13402573755\, ,
&
H''(r_+)&\simeq-2.50355915586\, .
\label{eq:loc-np-derivatives}
\end{align}
The algebraic reconstruction of the metric and the sign conditions ensuring
regularity are given in \eqref{eq:loc-np-D}--\eqref{eq:loc-np-signs}.

As a further check, we reconstruct the eleven-dimensional metric and
four-form and substitute them into the supergravity equations. As mentioned
in the main text, for the solution quoted above the Bianchi identity is
satisfied numerically to about $10^{-15}$, the four-form equation of motion
to about $10^{-9}$, and the Einstein equation to about $10^{-10}$.

\subsection{The six flux periods}
\label{subsec:loc-np-flux}
The integral generators, twisting class and six flux
coefficients are defined in \eqref{eq:loc-np-twist}--\eqref{eq:loc-np-flux-classes}.
These six quantities may be obtained directly from the endpoint data.
In terms of the functions $Q_i$ defined in
\eqref{eq:loc-np-Q-definitions}, define
\begin{align}\label{eq:loc-np-endpoint-period-definitions}
I_4
&=
\frac12\left[\frac{H'}{Q_1}\right]_{r_-}^{r_+}
-\frac14(r_+-r_-)\, ,
\qquad 
I_5
=
\frac12\left[\frac{H'}{Q_2}\right]_{r_-}^{r_+}
-\frac14(r_+-r_-)\, ,
\nonumber\\
I_6
&=
\frac12\left[\frac{H'}{Q_3}\right]_{r_-}^{r_+}\, ,
\nonumber\\
m_1
&=
\frac{H'(r_-)}{Q_3(r_-)}+2c_1\, ,
\qquad
m_2
=
\frac{H'(r_-)}{Q_3(r_-)}+2c_2\, ,
\nonumber\\
m_0
&=
H'(r_-)
\left[
\frac{1}{Q_1(r_-)}
+\frac{1}{Q_2(r_-)}
\right]
-r_-+2c_3\, .
\end{align}
For this numerical solution, these are
\begin{align}\label{eq:loc-np-unnormalized-periods}
I_4&=-0.18714538461432\, ,
\qquad\ \mskip2mu
I_5=-0.20585992307576\, ,
\nonumber\\
I_6&=-1.13637332757882\, ,
\qquad
m_0=-0.524007076920112\, ,
\nonumber\\
m_1&=3.78791109192922\, ,
\qquad \ \ \mskip2mu
m_2=3.78791109192922\, .
\end{align}

Restoring an overall length scale $L$ into the supergravity solution  we define
\begin{equation}\label{eq:loc-np-A-definition}
A
\equiv
8\pi^2\frac{L^3}{(2\pi\ell_p)^3}
=
\frac{L^3}{\pi\ell_p^3}\, .
\end{equation}
Denoting by $t=\frac{1}{4\pi}\int_{T^2}J_3$ the dimensionless area of the torus, with $J_3$ its K\"ahler form, the six magnetic
periods are then
\begin{align}\label{eq:loc-np-period-map}
N_0&=AtI_6\, ,
\qquad 
\mskip4mu  N_1=AI_4\, ,
\qquad 
\mskip17mu N_2=AI_5\, ,
\nonumber\\
M_0&=Am_0\, ,
\qquad
M_1=Atm_1\, ,
\qquad
M_2=Atm_2\, .
\end{align}
We choose the overall scale and torus area so that
$M_0=-28$ and $M_1=10$. This gives
\begin{align}\label{eq:loc-np-scale}
A
&=
-\frac{28}{m_0}
=
53.4343928417378\ldots\, ,
\nonumber\\
t
&=
\frac{10}{Am_1}
=
0.0494059601908478\ldots\, .
\end{align}
At the precision of the numerical data quoted above, the remaining four
periods agree with integers to approximately $3\times10^{-13}$. We thus
obtain the unshifted flux vector
\begin{equation}\label{eq:loc-np-unshifted}
(N_0,N_1,N_2; M_0,M_1,M_2)
=
(-3,-10,-11; -28,10,10)\, .
\end{equation}
The resulting fibre class, compatibility condition and opposite pole flux
are given in section~\ref{subsec:loc-np}.

\subsection{Central charge from endpoint and flux data}
\label{subsec:loc-np-central}
The Brown--Henneaux expression gives the warped volume with the
flat torus area factored out. Although written as a radial integral, this
quantity reduces to endpoint data, as explained below,
\begin{equation}\label{eq:loc-np-volume}
\widehat V_w^{(0)}
\equiv
\frac{1}{t}\int_{M_8}\me^{9\lambda}\vol_8
=
209.9828249266182\, .
\end{equation}
Thus, $V_w^{(0)}=t\widehat V_w^{(0)}$. Restoring the scale fixed in
\eqref{eq:loc-np-A-definition} and \eqref{eq:loc-np-scale}, the
Brown--Henneaux formula gives
\begin{equation}\label{eq:loc-np-csol}
c_{\mathrm{sol}}^{(0)}
=
\frac{3A^3t}{32\pi^4}\widehat V_w^{(0)}
=
1523.34243709687\ldots\, .
\end{equation}

The same central charge may instead be expressed entirely in terms of the
magnetic fluxes. Specializing the general formula
\eqref{eq:KE4-central-oriented} to
$KE_4=S^2_1\times S^2_2$, with
$c_1(\mathcal L)=2(\omega_1+\omega_2)$, gives
\begin{equation}\label{eq:loc-np-cloc-formula}
c_{\mathrm{loc}}^{(0)}
=
-6N_0N_1N_2
\left[
1+
\frac{4N_0^2(N_1+N_2)^2}{\Delta}
\right]\, ,
\end{equation}
where
\begin{align}\label{eq:loc-np-Delta}
\Delta
={}&
4N_1N_2M_1M_2
-\left(N_1M_2+N_2M_1-N_0M_0\right)^2
+
4N_0N_1N_2(M_1+M_2)\, .
\end{align}
For the flux vector \eqref{eq:loc-np-unshifted}
$\Delta=-68836$, 
and hence
\begin{equation}\label{eq:loc-np-cloc}
c_{\mathrm{loc}}^{(0)}
=
\frac{26215200}{17209}
=
1523.3424370968679\ldots\, .
\end{equation}
For this ansatz, the warped volume integral reduces to endpoint data because
$D=2H''-(Q_1+Q_2)$ and the $Q_i$ are fixed polynomials. Its equality with
the localized flux expression follows from the endpoint identity proved in
\cite{Gregory:2026nonpolynomial}. Thus localization expresses the same
endpoint quantity entirely in terms of global flux and topological data;
the existence of a regular interior solution is a separate question.

\subsection{Gravitational anomaly}
\label{subsec:loc-np-anomaly}
The gravitational anomaly follows immediately from
\eqref{eq:S2-anomaly-general}. For
$B_6=S^2_1\times S^2_2\times T^2$ one has
\begin{equation}
p_1(TB_6)=0\, ,
\qquad
c_1(\mathcal L)^2
=
8\,\omega_1\wedge\omega_2\, .
\end{equation}
Consequently only the $N_0\tau$ component of the fibre flux contributes,
and
\begin{equation}\label{eq:loc-np-anomaly}
c_L-c_R
=
-2N_0
=
6\, .
\end{equation}

\subsection{Exclusion of polynomial solution}\label{app:polynomial-elimination}

Here we prove the implication \eqref{eq:loc-np-poly-equal-flux} for regular compact
solutions of \eqref{eq:loc-np-master} satisfying
\eqref{eq:loc-np-boundary} and \eqref{eq:loc-np-signs}. In particular,
$Q_i>0$ on the closed interval and $H>0$ between two distinct simple zeros.
The radial translation setting $c_3=0$ preserves polynomiality of the solution.

First, every polynomial solution has degree at most three. If
$H=h_n r^n+\cdots$ with $n\geq4$, the coefficient of $r^{2n-2}$ in
$2HH''-(H')^2$ is
\begin{equation}\label{eq:poly-app-degree}
n(n-2)h_n^2\,.
\end{equation}
The remaining terms in \eqref{eq:loc-np-master} have degree at most $n+1$
and four, respectively, so cannot cancel this coefficient. Hence $h_n=0$,
a contradiction.

Now suppose that $M_1=M_2$. The endpoint period formulae
\eqref{eq:loc-np-endpoint-period-definitions} and the normalization
\eqref{eq:loc-np-period-map} give
\begin{equation}\label{eq:poly-app-equal-poles}
M_1-M_2=2At(c_1-c_2)\,.
\end{equation}
Since $A,t>0$, it follows that $c_1=c_2=\gamma$. If $d_1=d_2$, then $Q_1=Q_2$ and
$I_4=I_5$, which gives $N_1=N_2$ as required. It remains to exclude
$d_1\neq d_2$ for a regular compact polynomial solution.

Write
\begin{equation}\label{eq:poly-app-cubic}
H=ar^3+br^2+ur+v\,,
\end{equation}
allowing the leading coefficients to vanish. Substituting into
\eqref{eq:loc-np-master} gives the following five coefficient equations,
listed in descending powers of $r$ and with constant numerical factors
removed:
\begin{align}\label{eq:poly-app-coefficients}
0&=(a+2\gamma)(3a+2\gamma)\,,\nonumber\\
0&=(a+2\gamma)b+(a+\gamma)(d_1+d_2)\,,\nonumber\\
0&=(3a+4\gamma)u+2b(d_1+d_2)+2d_1d_2-8\gamma^2d_3\,,\nonumber\\
0&=(3a+2\gamma)v+(d_1+d_2)(u-4\gamma d_3)\,,\nonumber\\
0&=4(b+d_1+d_2)v-u^2-16d_1d_2d_3\,.
\end{align}

If $\gamma=0$, the first equation gives $a=0$. Positivity of $Q_1,Q_2$
requires $d_1<0$,  $d_2<0$, so the third and fourth equations imply
\begin{equation}\label{eq:poly-app-quadratic}
b=-\frac{d_1d_2}{d_1+d_2}>0\,,\qquad u=0\,.
\end{equation}
A convex quadratic is negative between any two distinct real
zeros, contradicting $H>0$. Thus, no regular compact polynomial solution
occurs on this locus.

For $\gamma\neq0$, the first equation in
\eqref{eq:poly-app-coefficients} gives $a=-2\gamma$ or
$a=-2\gamma/3$, i.e., in particular, the polynomial is genuinely cubic.
For $a=-2\gamma$, the second equation immediately gives $d_1+d_2=0$.
For $a=-2\gamma/3$, the second and third equations instead give
\begin{equation}\label{eq:poly-app-second-branch}
b=-\frac{d_1+d_2}{4}\,,\qquad
u=\frac{(d_1-d_2)^2+16\gamma^2d_3}{4\gamma}\,.
\end{equation}
The fourth equation then reduces to
$(d_1+d_2)(d_1-d_2)^2=0$. Under the assumption $d_1\neq d_2$, both
branches therefore require $d_2=-d_1$. Set $d_1=d\neq0$. The remaining
coefficient equations give
\begin{equation}\label{eq:poly-app-degenerate-data}
4\gamma^2d_3=d^2\,,\qquad Q_3=r^2-\frac{d^2}{\gamma^2}\,,
\end{equation}
and the two possible polynomials are
\begin{equation}\label{eq:poly-app-two-branches}
\begin{aligned}
H&=-2\gamma r^3+br^2-\frac{2d^2}{\gamma}r
&&\text{for }a=-2\gamma\,,\\
H&=-\frac{2\gamma}{3}r^3+\frac{2d^2}{\gamma}r+v
&&\text{for }a=-2\gamma/3\,.
\end{aligned}
\end{equation}
Here $b$ in the first branch and $v$ in the second remain free.

Neither branch has the required endpoint slopes. In the first,
$H'=2H/r-2\gamma Q_3$ for $r\neq0$. Since $Q_1(0)Q_2(0)=-16d^2<0$,
its zero at $r=0$ is not a regular endpoint. In the second,
$H'=-2\gamma Q_3$ identically. Thus, every admissible endpoint in either
branch obeys
\begin{equation}\label{eq:poly-app-slope-obstruction}
H'(r_\pm)=-2\gamma Q_3(r_\pm)\,.
\end{equation}
Because $Q_3(r_\pm)>0$ and $\gamma\neq0$, these slopes have the same
sign. A positive solution between simple zeros requires
$H'(r_-)>0$ and $H'(r_+)<0$, so this is impossible.

Consequently $d_1=d_2$ is necessary whenever a regular compact polynomial
solution has $M_1=M_2$, proving \eqref{eq:loc-np-poly-equal-flux}. The flux vector in \eqref{eq:loc-np-poly-benchmark} does not satisfy this condition and therefore has no regular compact polynomial representative within this ansatz. This covers both cubic branches and the lower degree cases.

\section{Dual flux periods}
\label{app:a0-dual-flux}

The scalar potential $a_0$ introduced via
\begin{equation}
\me^{3\lambda}f=\diff a_0
\end{equation}
is defined only up to an additive constant. In this appendix, we clarify how
this freedom is related to a condition imposed in
\cite{BenettiGenolini:2023yfe, BenettiGenolini:2023ndb}. Recall that
\begin{equation}
y=\frac12\left(\me^{3\lambda}\zeta-a_0\right)\, ,
\end{equation}
and that the closed four-form entering the equivariant completion of the
dual flux is
\begin{equation}
\Phi^{*F}_4
=
\me^{6\lambda}*F-a_0\me^{3\lambda}F\, .
\end{equation}
Under the constant shift
\begin{equation}
a_0\longrightarrow a_0+c\, ,
\end{equation}
one has
\begin{equation}
y\longrightarrow y-\frac{c}{2}\, ,
\qquad
\Phi^{*F}_4\longrightarrow
\Phi^{*F}_4-c\,\me^{3\lambda}F\, .
\end{equation}
Thus, defining the periods
\begin{equation}
Q_{*F}(\mathcal C_4)
\equiv
\int_{\mathcal C_4}
\left(\me^{6\lambda}*F-a_0\me^{3\lambda}F\right),
\qquad
Q_F(\mathcal C_4)
\equiv
\int_{\mathcal C_4}\me^{3\lambda}F\, ,
\end{equation}
their transformation is
\begin{equation}
Q_{*F}(\mathcal C_4)\longrightarrow
Q_{*F}(\mathcal C_4)-c\,Q_F(\mathcal C_4)\, .
\end{equation}

For an $S^4$ fibre carrying non-zero M5-brane charge,
$Q_F(S^4)\neq0$. The additive constant in $a_0$ may therefore always be
chosen so that
\begin{equation}
Q_{*F}(S^4)=0\, .
\end{equation}
This condition was imposed in
\cite{BenettiGenolini:2023yfe, BenettiGenolini:2023ndb}. From the present perspective, it fixes the
additive freedom in $a_0$, rather than constituting an additional
restriction on the supergravity solution, as was described in \cite{BenettiGenolini:2023yfe, BenettiGenolini:2023ndb}.

This may also be seen directly in the fixed point variables used in
\cite{BenettiGenolini:2023yfe, BenettiGenolini:2023ndb}. Localization of the magnetic flux through
the $S^4$ fibre is proportional to $y_+-y_-$, whereas the corresponding
period of $\Phi^{*F}_4$ is proportional to
\begin{equation}
y_+^2-y_-^2
=
(y_+-y_-)(y_++y_-)\, .
\end{equation}
For non-zero M5-brane charge one has $y_+\neq y_-$, so imposing the
vanishing dual period is equivalent to
\begin{equation}\label{yppym}
y_++y_-=0\, .
\end{equation}
Since a constant shift of $a_0$ translates both $y_+$ and $y_-$ by the
same amount, this is precisely a choice of origin for $y$. One can explicitly check that  the central charge in \cite{BenettiGenolini:2023yfe,BenettiGenolini:2023ndb} is independent of the additive constant in $a_0$, as it must be, and that the additional condition \eqref{yppym} imposed in those references is just a convenient gauge choice.  

More generally, a single additive constant can be used to fix one such
period. If several independent four-cycles carry magnetic flux, choosing a reference cycle for which ${Q_F(\mathcal C_4^{(0)})}\neq 0$, 
combinations such as
\begin{equation}
Q_{*F}(\mathcal C_4)
-
\frac{Q_F(\mathcal C_4)}
     {Q_F(\mathcal C_4^{(0)})}
Q_{*F}(\mathcal C_4^{(0)})
\end{equation}
are invariant under constant shifts of $a_0$ and contain the corresponding
gauge independent information.
\bibliographystyle{utphys} 
 \bibliography{AdS3M8_references}{}

\end{document}